\documentclass[twocolumn, twocolappendix]{aastex701}

\DeclareRobustCommand{\ion}[2]{\textup{#1\,\textsc{\lowercase{#2}}}}
\newcommand*\element[1][]{%
  \def\aa@element@tr{#1}%
  \aa@element
}
\usepackage{amsmath}
\usepackage{booktabs} 
\usepackage{tablefootnote}
\usepackage{lipsum}
\usepackage{threeparttable}
\usepackage{CJK}

\begin{document}
\begin{CJK*}{UTF8}{gbsn}

\title{The ALPINE-CRISTAL-JWST Survey:\\
Chemical Abundance Comparison Between the ISM and CGM of Main-Sequence Galaxies at $z=4-6$}

\suppressAffiliations
\correspondingauthor{Wuji Wang}
\email{wujiwang@ipac.caltech.edu}

\author[0000-0002-7964-6749]{Wuji Wang (王无忌)}
\affiliation{Caltech/IPAC, 1200 E. California Blvd. Pasadena, CA 91125, USA}
\email{}
\author[0000-0002-9382-9832]{Andreas L. Faisst}
\affiliation{Caltech/IPAC, 1200 E. California Blvd. Pasadena, CA 91125, USA}
\email{}
\author[0000-0002-4462-0709]{Kyle Finner}
\affiliation{Caltech/IPAC, 1200 E. California Blvd. Pasadena, CA 91125, USA}
\email{}
\author[0000-0002-3258-3672]{Livia Vallini}
\affiliation{INAF – Osservatorio di Astrofisica e Scienza dello Spazio di Bologna, Via Gobetti 93/3, 40129 Bologna, Italy}
\email{}
\author[0000-0002-7129-5761]{Andrea Pallottini}
\affiliation{Dipartimento di Fisica ``Enrico Fermi'', Universit\'{a} di Pisa, Largo Bruno Pontecorvo 3, Pisa I-56127, Italy}
\email{}
\author[0009-0007-1304-7771]{Enrico Veraldi}
\email{everaldi@sissa.it}
\affiliation{Scuola Internazionale Superiore Studi Avanzati (SISSA), Physics Area, Via Bonomea 265, 34136 Trieste, Italy}
\affiliation{Istituto Nazionale di Fisica Nucleare (INFN)-Sezione di Trieste, via Valerio 2, 34127 Trieste, Italy}
\affiliation{Istituto di Radioastronomia dell'Istituto Nazionale di Astrofisica (IRA-INAF), Via Gobetti 101, 40129 Bologna, Italy}

\author[0000-0001-5846-4404]{Bahram Mobasher}
\email{mobasher@ucr.edu}
\affiliation{Department of Physics and Astronomy, University of California, Riverside, 900 University Ave., Riverside, CA 92521, USA}

\author[0000-0001-8792-3091]{Yu-Heng Lin}
\email{ianlin@ipac.caltech.edu}
\affiliation{Caltech/IPAC, 1200 E. California Blvd. Pasadena, CA 91125, USA}

\author[0000-0002-2318-301X]{Giovanni Zamorani}
\affiliation{INAF – Osservatorio di Astrofisica e Scienza dello Spazio di Bologna, Via Gobetti 93/3, 40129 Bologna, Italy}
\email{}

\author[0000-0002-5877-379X]{Vicente Villanueva}
\affiliation{Departamento de Astronom{\'i}a, Universidad de Concepci{\'o}n, Barrio Universitario, Concepci{\'o}n, Chile}
\email{}

\author[0000-0002-3158-6820]{Sylvain Veilleux}
\affiliation{Department of Astronomy, University of Maryland, College Park, MD 20742 USA}
\email{}

\author[0000-0002-2645-679X]{Keerthi Vasan G.C.}
\affiliation{The Observatories of the Carnegie Institution for Science, 813 Santa Barbara Street, Pasadena, CA 91101, USA}
\email{}

\author[0000-0003-4891-0794]{Hannah \"Ubler}
\affiliation{Max-Planck-Institute f\"ur extratarrestrische Physik, Giessenbachstrasse 1, 85748 Garching, Germany}
\email{}

\author[0000-0002-0498-5041]{Akiyoshi Tsujita}
\email{tsujita@ioa.s.u-tokyo.ac.jp}
\affiliation{Institute of Astronomy, Graduate School of Science, The University of Tokyo, 2-21-1 Osawa, Mitaka, Tokyo 181-0015, Japan}

\author[0000-0002-7919-245X]{Kseniia Telikova}
\email{k.telikova@gmail.com}
\affiliation{Instituto de Estudios Astrof\'isicos, Facultad de Ingenier\'ia y Ciencias, Universidad Diego Portales, Av. Ej\'ercito Libertador 441, Santiago 8370191, Chile}

\author[[0000-0002-0000-6977]{John D. Silverman}
\email{}
\affiliation{Kavli Institute for the Physics and Mathematics of the Universe (Kavli IPMU, WPI), UTIAS, Tokyo Institutes for Advanced Study, University of Tokyo, Chiba, 277-8583, Japan}
\affiliation{Department of Astronomy, Graduate School of Science, The University of Tokyo, 7-3-1 Hongo, Bunkyo, Tokyo 113-0033, Japan}
\affiliation{Center for Data-Driven Discovery, Kavli IPMU (WPI), UTIAS, The University of Tokyo, Kashiwa, Chiba 277-8583, Japan}
\affiliation{Center for Astrophysical Sciences, Department of Physics \& Astronomy, Johns Hopkins University, Baltimore, MD 21218, USA}

\author[0000-0002-9948-3916]{Michael Romano}
\email{mromano@mpifr-bonn.mpg.de}
\affiliation{Max-Planck-Institut f\"ur Radioastronomie, Auf dem H\"ugel 69, 53121 Bonn, Germany}
\affiliation{INAF - Osservatorio Astronomico di Padova, Vicolo dell'Osservatorio 5, I-35122 Padova, Italy}

\author[0000-0003-1682-1148]{Monica Relano}
\affiliation{Dept. F\'isica Te\'{o}rica y del Cosmos, Campus de Fuentenueva, Edificio Mecenas, Universidad de Granada, E-18071, Granada, Spain}
\affiliation{Instituto Universitario Carlos I de Física Te\'{o}rica y Computacional, Universidad de Granada, 18071, Granada, Spain}
\email{}

\author[0000-0002-7412-647X]{Francesca Pozzi}
\affiliation{University of Bologna – Department of Physics and Astronomy “Augusto Righi” (DIFA), Via Gobetti 93/2, 40129 Bologna, Italy}
\affiliation{INAF – Osservatorio di Astrofisica e Scienza dello Spazio di Bologna, Via Gobetti 93/3, 40129 Bologna, Italy}

\email{}

\author[0000-0001-6652-1069]{Ambra Nanni}
\affiliation{National Centre for Nuclear Research, ul. Pasteura 7, 02-093 Warsaw, Poland}
\affiliation{INAF - Osservatorio astronomico d'Abruzzo, Via Maggini SNC, 64100, Teramo, Italy}
\email{}

\author[0000-0002-8136-8127]{Juan Molina}
\email{juan.molinato@uv.cl}
\affiliation{Instituto de F\'{i}sica y Astronom\'{i}a, Universidad de Valpara\'{i}so, Avda. Gran Breta\~{n}a 1111, Valpara\'{i}so, Chile}
\affiliation{Millenium Nucleus for Galaxies (MINGAL), Avda. Gran Breta\~{n}a 1111, Valpara\'{i}so, Chile}

\author[0009-0004-1270-2373]{Lun-Jun Liu}
\email{lliu@caltech.edu}
\affiliation{Physics Department, California Institute of Technology, 1200 E. California Blvd., Pasadena, CA, 91125 USA}

\author{Yuan Li}
\affiliation{Department of Physics and Astronomy and George P. and Cynthia Woods Mitchell Institute for Fundamental Physics and Astronomy, Texas A\&M University, 4242}
\email{}

\author[0000-0003-1041-7865]{Mahsa Kohandel}
\affiliation{Scuola Normale Superiore, Piazza dei Cavalieri 7, I-56126 Pisa, Italy}
\email{}

\author[0000-0002-6610-2048]{Anton M. Koekemoer}
\email{koekemoer@stsci.edu}
\affiliation{Space Telescope Science Institute, 3700 San Martin Drive, Baltimore, MD 21218, USA}

\author[0000-0003-4268-0393]{Hanae Inami}
\affiliation{Hiroshima Astrophysical Science Center, Hiroshima University, 1-3-1 Kagamiyama, Higashi-Hiroshima, Hiroshima 739-8526, Japan}
\email{}

\author[0009-0008-9801-2224]{Edo Ibar}
\affiliation{Instituto de F\'{i}sica y Astronom\'{i}a, Universidad de Valpara\'{i}so, Avda. Gran Breta\~{n}a 1111, Valpara\'{i}so, Chile}
\affiliation{Millenium Nucleus for Galaxies (MINGAL), Avda. Gran Breta\~{n}a 1111, Valpara\'{i}so, Chile}
\email{}

\author[0000-0002-2775-0595]{Rodrigo Herrera-Camus}
\affiliation{Departamento de Astronom\'ia, Universidad de Concepci\'on, Barrio Universitario, Concepci\'on, Chile}
\affiliation{Millenium Nucleus for Galaxies (MINGAL), Av. Ej\'ercito 441, Santiago 8370191, Chile}
\email{}

\author[0009-0003-3097-6733]{Ali Hadi}
\email{ahadi005@ucr.edu}
\affiliation{Department of Physics and Astronomy, University of California, Riverside, 900 University Ave, Riverside, CA 92521, USA}

\author[0009-0005-8932-7783]{Nicol Guti\'errez-Vera}
\affiliation{Departamento de Astronom\'ia, Universidad de Concepci\'on, Barrio Universitario, Concepci\'on, Chile}
\affiliation{Millenium Nucleus for Galaxies (MINGAL), Av. Ej\'ercito 441, Santiago 8370191, Chile}
\email{}

\author[0000-0002-9122-1700]{Michele Ginolfi}
\email{michele.ginolfi@unifi.it}
\affiliation{Universit\`a di Firenze, Dipartimento di Fisica e Astronomia, via G. Sansone 1, 50019 Sesto Fiorentino, Florence, Italy}
\affiliation{INAF -- Arcetri Astrophysical Observatory, Largo E. Fermi 5, I-50125, Florence, Italy}

\author[0000-0001-7201-5066]{Seiji Fujimoto}
\affiliation{David A. Dunlap Department of Astronomy and Astrophysics, University of Toronto, 50 St. George Street, Toronto, Ontario, M5S 3H4, Canada}
\affiliation{Dunlap Institute for Astronomy and Astrophysics, 50 St. George Street, Toronto, Ontario, M5S 3H4, Canada}
\email{}

\author[0000-0003-0348-2917]{Miroslava Dessauges-Zavadsky}
\affiliation{D\'epartement d'Astronomie, Universit\'e de Gen\`eve, Chemin Pegasi 51, 1290 Versoix, Switzerland}
\email{}

\author[0000-0001-9419-6355]{Ilse De Looze}
\affiliation{Sterrenkundig Observatorium, Ghent University, Krijgslaan 281 - S9, B-9000 Gent, Belgium}
\email{}

\author[0009-0007-7842-9930]{Poulomi Dam}
\affiliation{Dipartimento di Fisica e Astronomia Galileo Galilei Universit{\`a} degli Studi di Padova, Vicolo dell'Osservatorio 3, 35122 Padova, Italy}
\email{poulomi.dam@studenti.unipd.it}

\author[0000-0001-9759-4797]{Elisabete da Cunha} 
\affiliation{International Centre for Radio Astronomy Research (ICRAR), The University of Western Australia, M468, 35 Stirling Highway, Crawley, WA 6009, Australia}
\email{}

\author[0000-0003-0946-6176]{M\'ed\'eric Boquien}
\affil{Universit\'e C\^ote d'Azur, Observatoire de la C\^ote d'Azur, CNRS, Laboratoire Lagrange, 06000, Nice, France}
\email{}

\author[0000-0002-9508-3667]{Roberto J. Assef}
\affiliation{Instituto de Estudios Astrof\'isicos, Facultad de Ingenier\'ia y Ciencias, Universidad Diego Portales, Av. Ej\'ercito Libertador 441, Santiago 8370191, Chile}
\email{}

\author[0000-0002-6290-3198]{Manuel Aravena}
\affiliation{Instituto de Estudios Astrof\'isicos, Facultad de Ingenier\'ia y Ciencias, Universidad Diego Portales, Av. Ej\'ercito Libertador 441, Santiago 8370191, Chile}
\affiliation{Millenium Nucleus for Galaxies (MINGAL), Av. Ej\'ercito 441, Santiago 8370191, Chile}
\email{}

\author[0000-0002-1233-9998]{David B. Sanders}
\affiliation{Institute for Astronomy, University of Hawaii, 2680 Woodlawn Drive, Honolulu, HI 96822, USA}
\email{}

\collaboration{all}{(Affiliations can be found after the references)}



\begin{abstract}

Gaseous halos around galaxies play an important role in galaxy evolution. The exchange of metals from the interstellar medium (ISM) to the circumgalactic medium (CGM) are caused by the formation, feedback, and/or merging history of galaxies. We study the variation in chemical composition between the ISM ($\lesssim3\,$kpc) and CGM ($\sim5-10\,$kpc) for a sample of $M_{\star}>10^{9.5}\,M_{\odot}$ main-sequence galaxies at $4<z<6$ with both JWST and ALMA observations. Using JWST/NIRSpec integral field spectroscopy, we derive the optical line ratios from the ISM and the CGM for our sample focusing on the typical optical lines used for metallicity studies. Our comparison shows that the ISM and the CGM have similar chemical abundances. This indicates that the CGM of these typical $4<z<6$ galaxies is enriched to the level of their ISM in the early universe. Using statistical tests, we find that some of the line ratios show marginal differences between the ISM and CGM. Combined with \texttt{Cloudy} modeling, our results suggest that a difference in ionization level (higher for the ISM) is the dominant reason for the observed ratio difference of oxygen line ratios. There is also indication of a deficit in the nitrogen abundance with respect to oxygen in the CGM, which suggests a delay in redistribution of secondary nitrogen. Finally, an enhanced $F_{[\ion{C}{ii}]\rm 158\mu m}/F_{\rm H\alpha}$ ratio is observed in the CGM, suggesting that feedback and/or mergers play a key role in metal mixing. 

\end{abstract}



\section{Introduction} \label{sec:intro}

An important aspect of the study of galaxy formation and evolution is deciphering the interplay between their stars, interstellar medium (ISM), and their surrounding diffuse gaseous halos. The latter so-called circumgalactic medium \citep[CGM,][]{Tumlinson_2017} plays an important role in the buildup of galaxies by acting as a gas reservoir that feeds star formation. 
The CGM consists of pristine material accreted from larger galactic scales and gas expelled by feedback processes from supermassive black hole (SMBH) activity and stellar evolution \citep[e.g.,][]{Veilleux_2020,WangXin_2022,LuSD_2022}. These phenomena chemically enrich the circumgalactic and even intergalactic media with productions from the ISM.

Ejected gas that fails to escape the gravitational potential may be recycled, a key scenario that governs subsequent star formation and enrichment \citep[e.g.,][]{Oppenheimer_2010,Ford_2014,Ubler_2014,AnglesAlcazar_2017,Fraternali_2017}. Therefore, it is important to study the metal distribution and enrichment, the ionization properties, and dynamics throughout the galactic halos for a comprehensive understanding of the subsequent galaxy growth and evolution. Observational studies of the (chemical) properties of the CGM are often conducted through detection of emission lines \citep[e.g., in the surrounding of powerful quasars,][]{FAB_2019,Wang_2021_4C,Wang_2023_3D,Zhang_shiwu_2023} and absorption lines against bright background objects \citep[e.g., at 100$\,$kpc scales,][]{Churchill_2012, Kacprzak_2019, Mendez-Hernandez_2022}. Despite these efforts, a detailed understanding of the metal enrichment of the CGM at early stages of galaxy evolution is still lacking.

To understand the buildup of stellar masses and the evolution of galaxy structures in the early universe, a comprehensive study of the gaseous media surrounding the galaxy is needed soon after the epoch of reionization \citep[EoR, $z\gtrsim6$,][]{Zaroubi_2013EoR}. This is challenging due to limitation in sensitivity and wavelength coverage in high-$z$ observations, {\it i.e.} it is difficult to obtain the optical emission lines which serve as metallicity tracers \citep[e.g.,][]{Kewley_2002,Maiolino_2008,Steidel_2014}.
The advent of the James Webb Space Telescope (JWST) enables, for the first time, detailed studies of the galactic structure and CGM at high redshifts. Critically, JWST provides high sensitivity and high (spatial) resolution data at previously inaccessible rest-frame optical wavelengths for probing early metal enrichment at $z>3$ \citep[e.g.,][]{Nakajima_2023,Curti_2024,Morishita_2024,Vallini_2024,Venturi_2024,ZhangYechi_2025,Ji_Xihan_2025,Faisst_2025b}. In addition, multi-wavelength observations capturing the light of stars, dust, and gas at various spatial scales are crucial to establish a panoramic view of galaxy evolution. Efforts have also been made in using far-infrared (FIR) emission lines for studying metallicity at $z\sim3-6$ using primarily observations from ALMA \citep[e.g.,][]{Nagao_2012,Cunningham_2020,Vallini_2024,Harikane_2025}. Furthermore, combining FIR and optical emission lines, one can constrain gas density, metallicity, and the ionization parameter\citep[e.g.,][]{Vallini_2025}.

Studying the redshift range $z=4-6$ ($1.5-0.9\,$Gyr after the Big Bang) provides the opportunity to shed light on the period of galaxy evolution that links primordial galaxy formation during the EoR to the era of cosmic noon  \citep[e.g.,][]{Zaroubi_2013EoR,Madau_2014}. Multi-waveband data, tracing gas, stars, and dust are essential for such studies. For example, the large ALMA programs ALPINE and CRISTAL focus on this transition epoch offering insights on the main-sequence galaxies \citep[][]{LeFevre_2020,Faisst_2020,Bethermin_2020,Herrera-Camus_CRISTAL}. The ALPINE-CRISTAL-JWST survey \citep{Faisst_ALPINE_JWST,Fujimoto_ALPINE_JWST} targets galaxies in the COSMOS field from these ALMA large programs, thereby combining resolution matched, $1-2\,$kpc, multi-wavelength datasets from UV to FIR \citep[][]{Scoville_2007,LeFevre_2020,Herrera-Camus_CRISTAL,Franco_2025}. These rich datasets provide the knowledge of young and mature stellar populations, ionized and neutral gas, and dust for a comprehensive understanding of galaxies evolution.
Importantly, with the integral field spectrography on JWST (covering a $3\arcsec\times\arcsec3$ field of view, we can, for the first time, probe the CGM around these $z\sim5$ galaxies at larger physical scales \citep[$10-20\,{\rm kpc}$;][]{Faisst_ALPINE_JWST} --- in addition to studies of metallicity gradients that are usually limited to the ISM and galactic disks out to $\sim4\,$kpc \citep[e.g.,][]{WangXin_2022,Curti_2024,Venturi_2024,Ju_2025,Fujimoto_ALPINE_JWST,LLee_ALPINE_JWST}.

In this paper, we use JWST's IFU capabilities the probe the chemical properties farther into the CGM by stacking CGM pixel spectra. We compare, for the first time, the chemical composition difference between the ISM and CGM, at $z\sim5$, through emission line ratios in both optical and FIR using combined JWST/NIRSpec IFU and ALMA data. In Sect. \ref{sec:sample}, we briefly introduce the ALPINE-CRISTAL-JWST sample and the data used in this paper. Leveraging also the NIRCam imaging, we define the CGM to be out to $5-10\,$kpc by masking the central $\sim3\,$kpc ISM region (Sect. \ref{subsec:def_CGM}). We describe our spectral analysis in Sect. \ref{subsec:ana_specfit} and present the derived line ratios in Sect. \ref{sec:results}. In Sect. \ref{sec:discussion}, we discuss possible explanations of the observed line ratio differences. For this paper, we assume a flat Lambda-CDM cosmology with $H_{0} = 70\, \rm{km\,s^{-1}\,Mpc^{-1}}$ and $\Omega_{m}=0.3$. Following this cosmology, $\rm{1\,arcsec=5.6-6.8\, kpc}$ at the redshift range being studied in this work, $z=4-6$.

\section{Sample and observations}\label{sec:sample}

\subsection{ALPINE-CRISTAL-JWST sample}

We study the gas-phase chemical abundances in a sample of 18 star-forming main-sequence galaxies at $z=4-6$ as part of the ALPINE-CRISTAL-JWST sample \citep{Faisst_ALPINE_JWST,Fujimoto_ALPINE_JWST}, which leverages primarily on the ionized gas emission observed by JWST/NIRSpec IFU data (\#3045, PI: Faisst). The summary of the JWST survey, the sample description, and the first measurements are presented in \citet{Faisst_ALPINE_JWST}. In this paper, we analyze the same sample which is briefly summarized here. The sample is based on galaxies with [\ion{C}{ii}]158$\rm \mu m$ detections from the ALMA-ALPINE survey \citep{LeFevre_2020,Bethermin_2020,Faisst_2020}, from which we select a subsample of galaxies that have high-angular kpc-scale resolution ALMA observations from the CRISTAL-ALMA follow-up program \citep[][]{Herrera-Camus_CRISTAL} and rest-frame optical JWST/NIRCam imaging from the COSMOS-Web survey  \citep[][]{Casey_2023}. All 18 targets have stellar masses $M_{\star} = 10^{9.5} - 10^{11}\,\rm M_{\odot}$, star formation rate ${\rm SFR}>10\,\rm M_{\odot}\,yr^{-1}$, and bright [\ion{C}{ii}] emissions \citep[$>10^{8}\,L_{\odot}$, e.g.,][]{Faisst_2020,Bethermin_2020,Dessauges-Zavadsky_2020,Schaerer_2020,Ikeda_2025,Herrera-Camus_CRISTAL}. We follow \citet{Romano_2021} for the classification of mergers \citep[see also][for a summary]{Faisst_ALPINE_JWST}. The galaxies in this sample are metal-enriched at $10-70\%$ solar metal abundance.  The metallicity has been measured in relative oxygen to hydrogen abundance to hydrogen, $Z_{\rm neb}=12+\log(\rm O/H)=7.7-8.5$, using the same NIRSpec IFU data \citep[see][]{Faisst_2025b}\footnote{$Z_{\odot}\equiv\rm 12+\log(O/H)=8.69$ \citep[][]{Asplund_2009_Zsun}}. Throughout the paper, we assume the fiducial gas-phase metallicity of $Z_{\rm neb}=0.5Z_{\odot}$ for the sample \citep[see also][]{Faisst_ALPINE_JWST}. This provides a representative sample of massive galaxies \citep[$\sim300\,\rm Myr$,][]{Faisst_2020,Tsujita_ALPINE_dust} for studying spatial enrichment around the time when the universe was $\sim0.9-1.5\,\rm Gyr$ old.

\subsection{JWST optical photometric and spectroscopic observations}

The NIRSpec/IFU data were observed through the program GO-3045. The disperser/filter G235M/F170LP (1.66$-$$3.07\,\rm\mu m$) and G395M/F290LP (2.87$-$$5.10\,\rm\mu m$) were set up ensuring a line coverage from [\ion{O}{ii}]$\lambda\lambda3726,3729$ to [\ion{S}{ii}]$\lambda\lambda6716,6731$. These medium resolution gratings offer a spectral resolution of $R\sim1000$ or $\Delta v \approx100-190\,\rm km\,s^{-1}$. For one galaxy, DC-842313, observations in high spectral resolution mode ($R\sim2700$ and $\Delta v \approx38-70\,\rm km\,s^{-1}$) were carried out under GO-4265 \citep[PI: J. Gonz\'{a}lez-Lopez;][]{Solimano_2025} with G395H/F170LP. The data processing is described in detail in \citet{Fujimoto_ALPINE_JWST}. The data cubes used in this work are sampled to have a pixel scale of $0.1\arcsec \, \rm pix^{-1}$. The typical spatial resolution is $\sim0.20\arcsec$. The \textit{JWST} data described here may be obtained from the MAST archive at
\dataset[doi:10.17909/cqds-qc81]{http://dx.doi.org/10.17909/cqds-qc81}.

In addition, we use the ancillary NIRCam data as part of the COSMOS-Web survey \citep[GO-1727, PIs: J. Kartaltepe \& C. Casey,][]{Casey_2023,Shuntov_2025_cosmos,Franco_2025}. The imaging data are more sensitive to continuum emission compared to NIRSpec IFU spectroscopy. For one galaxy (DC-417567), which is not covered by COSMOS-Web, we include its HST/WFC3 F125W imaging from CANDELS \citep[][]{Grogin_2011,Koekemoer_2011}. It has been known that the astrometry of NIRSpec IFU observations has an offset of $\sim 0.1-0.2\arcsec$ \citep[e.g., due to uncertainties in the guidestar positions used to acquire the observations,][]{Faisst_ALPINE_JWST,Wang_2024_4C,Wang_2025_HzRGs}. We corrected this offset to a few tens of milli-arcseconds by matching the centroid of the continuum image to the centroid of a synthetic image created by collapsing the IFU cube over a similar wavelength \citep[see][]{Fujimoto_ALPINE_JWST}. 

\subsection{ALMA FIR observations}
We also utilize ALMA observations of redshifted [\ion{C}{ii}]  in this work to study the carbon abundance in the CGM (Sect. \ref{subsec:C2dsut}). While the ALPINE-ALMA program provides these data \citep{Bethermin_2020}, we combined these data with the higher resolution observations from the CRISTAL-ALMA survey, which covers 15 targets in our sample \citep[\#2021.1.00280.L, PI: R. Herrera-Camus,][]{Herrera-Camus_CRISTAL}. The remaining three targets are observed under the programs \#2022.1.01118.S (PI: M. B\'ethermin) and \#2019.1.00226.S (PI: E. Ibar). All the high resolution ALMA observations were carried out in similar configurations where a typical resolution of $\sim0.3-0.4\arcsec$ was achieved. This is significantly better than previously available ALPINE data ($0.7-1.0\arcsec$) allowing for a spatial comparison to the JWST imaging and spectroscopic observations. In this work, we adopted the data cubes produced by combining both the high and low resolution observations. We used the cubes with {\it Briggs} weighting of $\texttt{robust}=+0.5$ in cleaning to avoid negative effects from side lobes that may be caused by using {\it Natural} weighting \citep[][]{Czekala_2021}. The data processing of these ALMA data can be found in \citet{Bethermin_2023} and \citet{Devereaux_2024}, respectively. The median synthesized beam size for the sample is $\theta_{\rm MAJ}\times\theta_{\rm MIN}\sim 0.36\arcsec\times0.30\arcsec$. The largest recoverable angular scale is $\gtrsim5\arcsec$ for the ALMA dataset which is larger than the focus of this work.

\begin{figure*}
    \centering
    \includegraphics[width=\textwidth,clip]{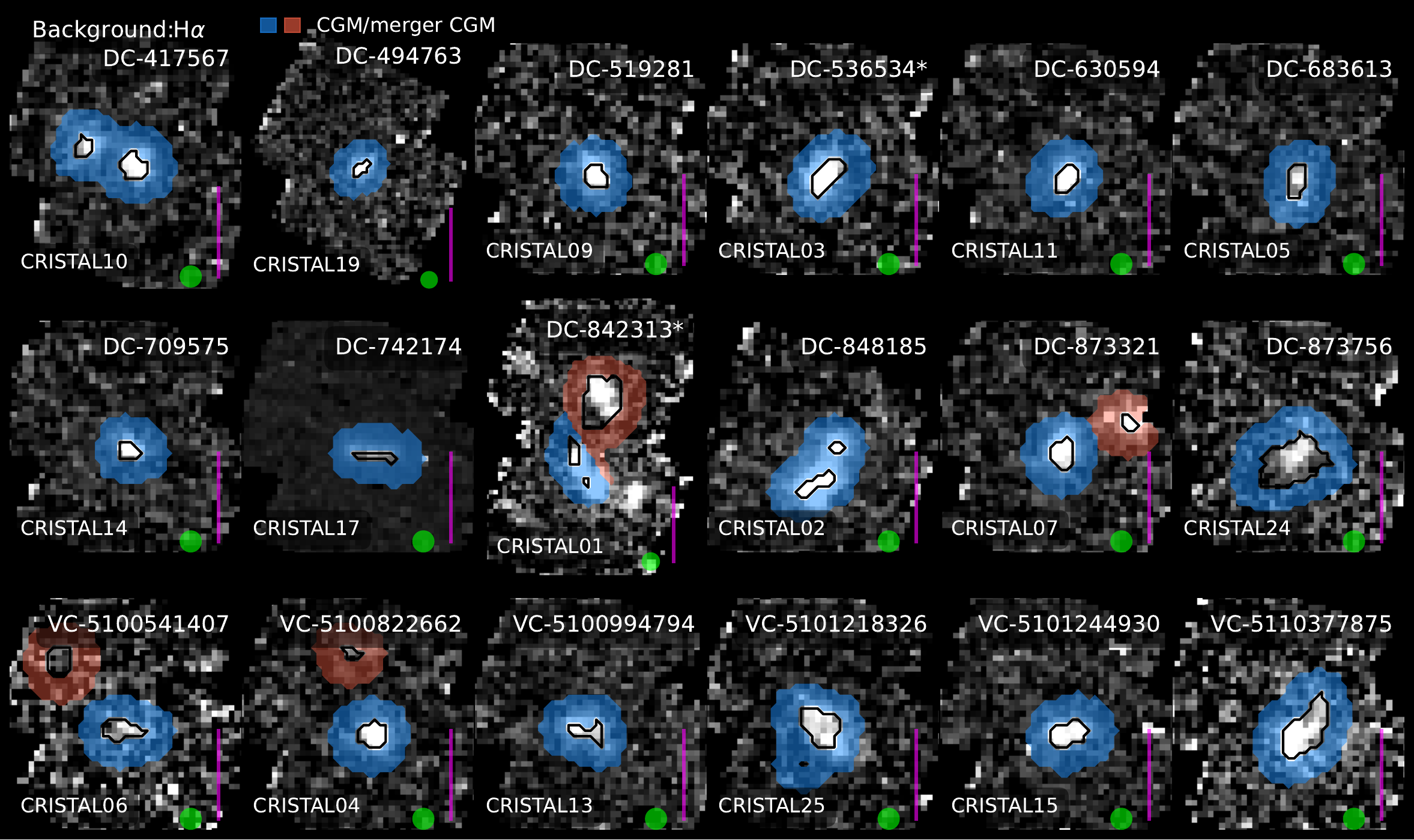}
    \caption{ISM and CGM masks overlaid on H$\alpha$ maps for our sample targets. The CGM masks are shaded blue, except for mergers, where we shade the companion CGM red. The outer boundary of the ISM regions are marked by the black contours. The pink line indicates the length of 10~kpc. The green circle with $r=0.2\arcsec$ represents the resolution element size \citep[][]{Fujimoto_ALPINE_JWST}. The CRISTAL ID number of the galaxies are given at the bottom left corner. Note that VC-5110377875 is not in the CRISTAL sample. We use asterisk to mark the galaxies host AGN \citep[see also][for other AGN candidates]{Ren_AGN}.}
    \label{fig:mask_pre}
\end{figure*}

\section{Analysis}\label{sec:analysis}

\subsection{Definition of ISM and CGM}\label{subsec:def_CGM}

The goal of this paper is to study the gas-phase chemical abundance differences in the ISM and CGM around typical main-sequence galaxies at $z\sim5$. Given the field of view (FoV) of the NIRSpec/IFU data, we can spatially cover out to $\sim20\,$kpc with an effective resolution of $\sim0.15\arcsec$.
This is a factor of $\sim 2-3$ smaller than the estimated virial radius of a galaxy with $M_{\star}=10^{10-10.5}\,M_{\odot}$ at $z=5$, which is $r_{\rm vir}\approx50-70\,{\rm kpc}$ \citep[e.g., ][]{Behroozi_2010,Behroozi_2013, Dekel_2013,Shibuya_2015,Tumlinson_2017}.
The virial radius is conventionally thought of as the outer boundary of the CGM \citep{Tumlinson_2017}. The interface between the ISM and the CGM can be estimated to be at $\sim5-7\,{\rm kpc}$ 
\citep[e.g.,][]{Nielsen_2024}.
In this work, we therefore focus on the intersection or interface between the ISM and the inner CGM -- an important regime where the chemical exchange between the ISM and outer CGM happens.

\begin{figure*}
    \centering
    \includegraphics[width=0.8\textwidth,clip]{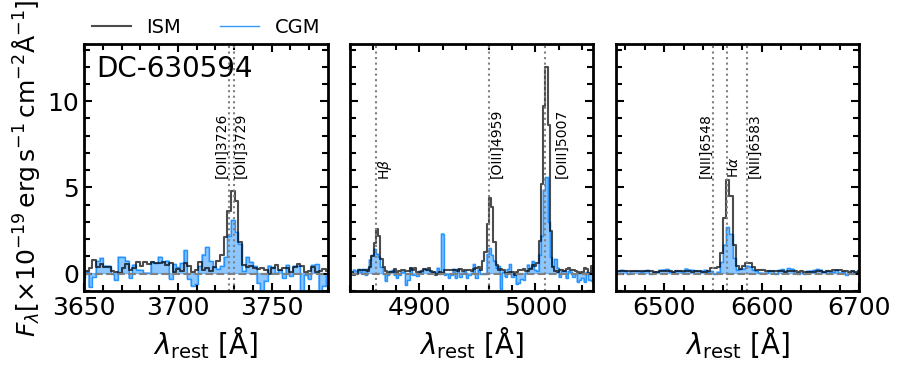}
    \caption{Example spectra extracted from the ISM and CGM of DC-630594. We mark the emission lines focused in this work.}
    \label{fig:example_sp}
\end{figure*}
\paragraph{The ISM region~}~
We use the NIRCam imaging of our sample, from the COSMOS-Web survey, as a proxy of the distribution of stars, thus the ISM of the galaxies. We define the ISM to be the region enclosed within the $9\sigma$ continuum level contour of the NIRCam image, PSF-matched to the NIRSpec IFU data. The $1\sigma$ noise level is estimated from source-free regions on the image. All the NIRCam images were obtained from COSMOS-Web and have the similar depth \citep[$\sim28$ AB magnitude,][]{Casey_2023,Shuntov_2025_cosmos}. The extent of the ISM has a median of $\sim3$ kpc, in good agreement with the expected extent of the ISM measured from the surface brightness radial profile \citep{Nielsen_2024} and typical  ``galaxy sizes'' at these redshifts \citep[two times the effective radius\footnote{For S\'ersic profile with $n=1$, $86\%$ of the total stellar light falls into two times the effective radius.},][]{vanderwel_2014,Martorano_2024}. This also agrees with the FIR continuum size of $\sim3\,$kpc measured for ALPINE galaxies \citep[][]{Pozzi_2024}. Our ISM region size is similar to or larger than one resolution element of NIRSpec IFU and ALMA. The ISM region of DC-742174 is compact due to its continuum morphology under our definition with NIRCam imaging. As a test, we relaxed the outer ISM edge to the $5\sigma$ flux contour of F444W and found that it does not impact our main results. Although a detailed morphology study is beyond the scope of this work, we conducted a \texttt{galight} fit on the F444W images for a manually selected subsample with no significant clumpiness to validate our ISM definition. Using a S\'ersic model with $n=1$, we found that the median $R_{e, \rm F444W} = 0.21\arcsec$ ($\sim1.4\,$kpc) from the fit is comparable to the half size of our flux based ISM.
Note that the choice of the continuum level for the definition of the ISM region does not have a significant impact on the following results because the light emission is dominated by the compact inner regions of the ISM and not its outskirts.

We use the observations in either the F277W and F444W filter, both covering the rest-frame optical emission from the bulk of evolved stellar populations to derive the extent of the ISM. Depending on the redshift of the galaxies, we choose the filter that avoids contamination by strong emission lines (H$\alpha$ and [\ion{O}{iii}]$\lambda5007$). For DC-417567 with no JWST images available, the HST/WFC3 F125W image is used \citep[e.g.,][]{Koekemoer_2011}. We convolve the NIRCam images with a $0.160\arcsec$ (F277W) or $0.135\arcsec$ (F444W) Gaussian kernel to account for the resolution difference between NIRCam and NIRSpec IFU observations.
To obtain the NIRCam PSF, we fit a 2D Gaussian profile to stars observed in the COSMOS-Web field in the respective filter.
To estimate the NIRSpec IFU PSF, we use the NIRSpec IFU observation of a calibrator star, 2MASS J17571324, and obtain a synthetic image by convolving the cube with the respective NIRCam filter transmission curve (JWST-GO-03399, PI: M. Perrin, A. Vayner private communication). We then perform a two-dimensional Gaussian fit to the convolved image to obtain the PSF size. The FWHM of the NIRSpec IFU PSF are in the range of $0.18\arcsec-0.21\arcsec$. This is consistent with the independent measurement, \citet{Fujimoto_ALPINE_JWST}, of the same dataset using \texttt{WebbPSF} \citep[][]{Perrin_2014} and literature studies \citep[e.g.,][]{Vayner_2024_q3d}. 
We note that the PSF shape in either case is not purely Gaussian, but a Gaussian approximation is accurate enough for our purpose ($\gtrsim80\%$ flux being taken into account).

\paragraph{The CGM region~}~
It is difficult to robustly define the CGM region observationally given that no clear common definition has been established \citep[][]{Tumlinson_2017}. This is particularly true at the boundary with the ISM to the CGM. The extended [\ion{C}{ii}] halo may trace the diffuse circumgalactic cold gas. For ALPINE-CRISTAL-JWST galaxies, thanks to the comprehensive baseline coverage of small and large spatial scales, \citet{Ikeda_2025} showed that the [\ion{C}{ii}] radial surface brightness profile does not exhibit a clear break out to the detection limit at $\sim15\,$kpc \citep[see also][for an example of extended halo detection from short baseline]{Akins_2022}. The [\ion{C}{ii}] halo may suggest an already carbon-enriched CGM, which could bias the study of abundances if focused on the [\ion{C}{ii}] halo alone. Motivated by the ISM region defined above, we therefore define the CGM region as the area enclosed by the expanded ISM mask excluding the original ISM.

In detail, we first smooth the ISM mask using the \texttt{binary\_dilation} function from \texttt{SciPy} Python package and then use a circular top-hat convolution kernel with a size of $0.5\arcsec$ to expand it. Under this treatment, we preserve the morphology of the ISM boundary when defining the CGM outer boundary, which is important as the ISM is not symmetric and spherical.
For known mergers with spatially separated components \citep[][]{Romano_2021}, we divide the CGM mask into two regions by applying a cut perpendicular to the line connecting the components of the merging systems. For DC-842313, we manually separate the halo between the eastern source and the western AGN \citep[{\it i.e.} excluding the known AGN dominating region in the south-west, see also][]{Solimano_2025}. This method produces CGM regions out to $r\sim10\,$kpc, $2-3$ times larger than the ISM regions and consistent with the observed ratio between UV and [\ion{C}{ii}] radii \citep[$r_{\rm UV}/r_{[\rm \ion{C}{ii}]}\sim2-3$,][]{Fujimoto_2020,Ikeda_2025}.
As mentioned above, this scale covers the important {\it interface} between the ISM and CGM. Note that our choice of the kernel size is an optimal trade-off between maximizing the CGM area and minimizing the inclusion of background noise. However, we found that our main conclusion of this work does not depend on the size of the CGM.
Representations of the ISM and CGM masks for our sample are shown in Fig.~\ref{fig:mask_pre}.

As a precaution, we tested whether setting the outer CGM boundary as either the $3\sigma$ detection limit of [\ion{C}{ii}] halo or the circular aperture with the radius of $4\,r_{[\rm \ion{C}{ii}]}$ \citep[][]{Ikeda_2025} would impact the results and found that neither of these two had a significant impact. The CGM size of these choice are all relatively comparable. Specifically, the smoothing kernel size in our current treatment (and the $4\,r_{[\rm \ion{C}{ii}]}$) is motivated by the depth of the ALMA observation {\it i.e.} the $3\sigma$ detection limit of [\ion{C}{ii}]. To avoid the bias towards the enriched diffuse halo, we do not use the [\ion{C}{ii}] to define the CGM.

\begin{figure*}
    \centering
    \includegraphics[width=\textwidth,clip]{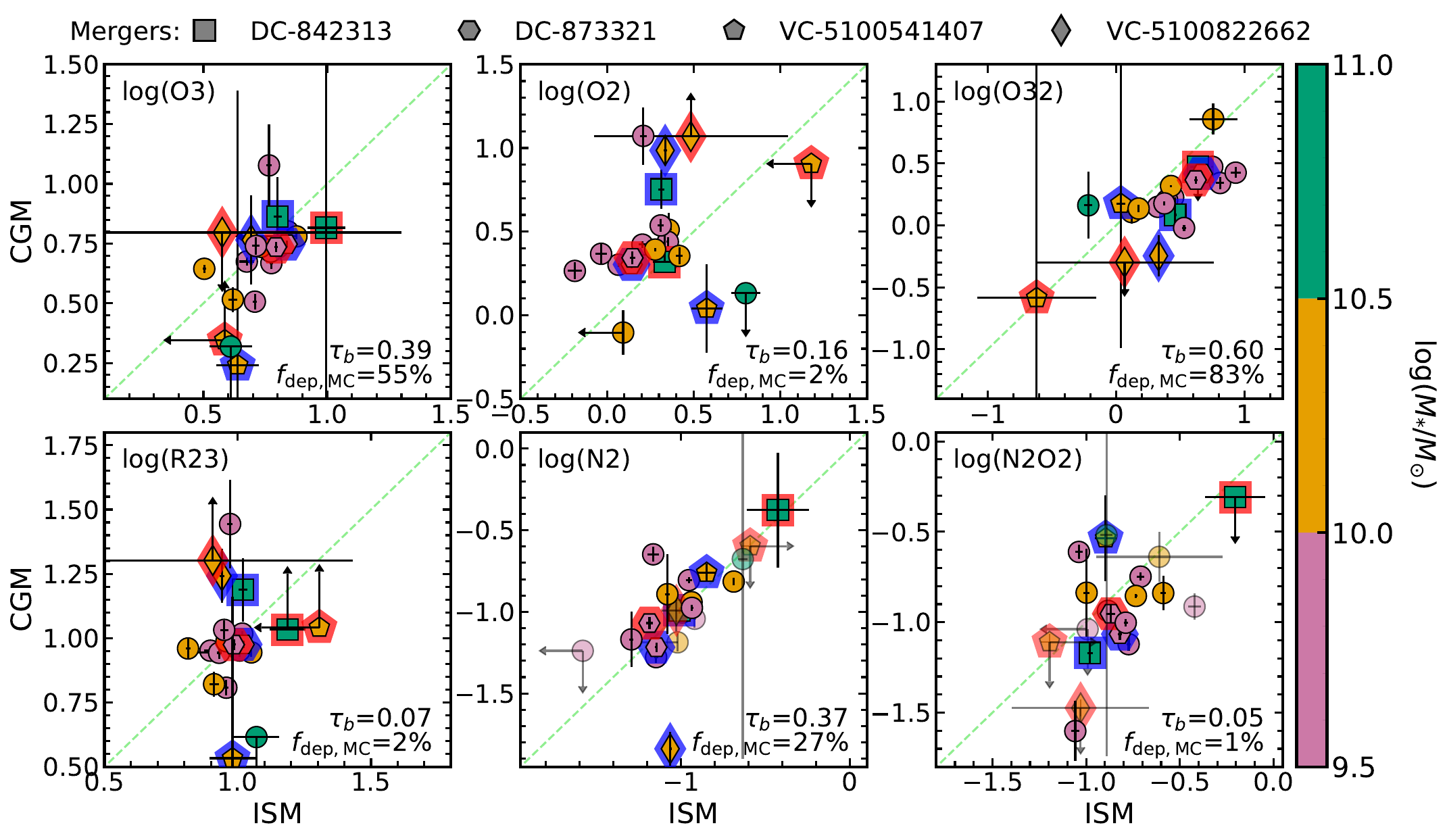}
    \caption{ISM versus CGM line ratios in logarithmic scales of our sample color-coded by $\log (M_{\star}/M_{\odot})$ in three bins. For mergers (see marker legend at top of figure), each ISM measure is linked to its CGM as described in Sect. \ref{subsec:def_CGM}. To distinguish between different merger components, we use blue and red edge colors for the merger symbols, corresponding to Fig.~\ref{fig:mask_pre}. We mark the one-to-one relation with a dashed green line. The targets in both log(N2) and log(N2O2) panels with S/N$_{[\ion{N}{ii}]\lambda6583}<5$ are shown in partial transparency (Sect. \ref{subsec:disc_n2o2}). We show the Kendall's $\tau$ coefficient and fraction of rejection based on MC sampling, {\it i.e.} the fraction of MC iterations where the probability of the ISM and CGM ratios are correlated is high at the bottom right ($f_{\rm dep, MC}$ $=f_{p<0.05}$, see text).
    }
    \label{fig:sample_lineratio}
\end{figure*}

\subsection{Spectral fitting}\label{subsec:ana_specfit}

We extract the total 1D spectra for the ISM and CGM masked regions identified in Sect. \ref{subsec:def_CGM}. We focus on strong optical emission lines, specifically, [\ion{O}{ii}]$\lambda\lambda3726,3729$, H$\beta$, [\ion{O}{iii}]$\lambda\lambda4959,5007$, [\ion{N}{ii}]$\lambda\lambda6548,6583$, and H$\alpha$, which indicate the ionization and metal abundance of the gas-phase medium \citep[][]{Maiolino_2008,Sanders_2024_Te}.  In Fig. \ref{fig:example_sp}, we show an example of the ISM and CGM spectra for DC-630594. Visual inspection suggests that the flux ratio of [\ion{O}{iii}] to  [\ion{O}{ii}] could be higher in the ISM. 
To derive robust total line fluxes, we perform Gaussian fits to each line using  \texttt{q3dfit} \citep[][]{Rupke_2023q3dfit}\footnote{\url{https://q3dfit.readthedocs.io/stable/}}.
We fit the lines in each of the two gratings separately as the sensitivity and resolution of the gratings can vary. For each grating, we fix the line redshifts and widths \citep[][]{Wylezalek_2022,Rupke_2023q3dfit}.
For all targets, [\ion{O}{ii}]$\lambda\lambda3726,3729$ are captured by G235M/F170LP and emission lines redward of [\ion{N}{ii}]$\lambda6548$ are observed in G395M/F290LP.
We note that for three sources at $5<z<5.3$ the H$\beta$ and [\ion{O}{iii}]$\lambda\lambda4959,5007$ lines fall in both the G235M and G395M gratings. Given the higher S/N of G395M, we adopt the results of the G395M grating for the following analysis. Comparing the fitted line fluxes from both gratings, we find that the ones from G395M are higher by $\sim6\%$. This difference in flux is minimal compared to the measurement uncertainties. The impact of this on extinction correction is also smaller than the uncertainties given that we study line pairs close in wavelength. The only ratio that might be influenced is $F_{[\ion{N}{ii}]\lambda6583}/F_{\rm[\ion{O}{ii}]\lambda3728}$, which we further discuss in Sect. \ref{subsec:disc_n2o2}. Nevertheless, our main conclusion is not affected by this flux difference across gratings, hence we did not take this into account in the following.
Since we only focus here on the total line flux (a dynamic analysis is beyond the scope of this paper) we do not distinguish different kinematic components and only fit a single Gaussian to the lines. If two Gaussian components are required, we summed the fluxes for line ratio calculation  \citep[][]{Solimano_2025}.

At medium spectral resolution ($R\sim1000$), we fix the flux ratio of the [\ion{O}{ii}]$\lambda\lambda3726,3729$ doublet to unity and refer to the total flux as [\ion{O}{ii}]$\lambda3728$. We further fix the doublet line ratio of $F_{\rm [\ion{O}{iii}]\lambda4959}/F_{\rm [\ion{O}{iii}]\lambda5007}$ and $F_{[\ion{N}{ii}]\lambda6548}/F_{[\ion{N}{ii}]\lambda6583}$ to one third. The resulting line fluxes are corrected for dust extinction using the observed Balmer decrement $F_{\rm H\alpha}/F_{\rm H\beta}$ and by applying a case B ratio of $2.86$ \citep{Cardelli_1989} and a Calzetti attenuation curve \citep{Calzetti_2000}. For the stellar-to-nebular dust attenuation ratio, we use $f=0.56$ following \citet{Tsujita_ALPINE_dust} for our sample. We note that the canonical case B recombination assumes $T_e=10^{4}\,$K and $n_{e}=10^{2}\,\rm cm^{-3}$ and may not be applicable universally \citep[e.g.,][]{Pirzkal_2024,Faisst_ALPINE_JWST,scarlata_2024}. In our case, the observed ratios in ISM are greater than $2.86$. However, the measurements in the CGM show a larger scatter. This may be due to different gas properties in the CGM, a wide variety of dust attenuation scenarios, or spread in dust attenuation values. For the observed ratio $F_{\rm H\alpha}/F_{\rm H\beta}<2.86$, we assume negligible dust attenuation and do not perform any correction. Most line ratios used in this work follow traditional line pairs where they are closer in wavelength thus are less affected by dust extinction. 

To obtain the uncertainties of the fitted quantities, a Monte Carlo (MC) procedure was adopted where we generated $100$ realizations of the data cubes by randomly sampling the noise distribution of each volume pixel \citep[a similar approach as used in][]{Faisst_ALPINE_JWST}. The noise includes instrumental effects and sky background emission and is assumed to follow a Gaussian distribution \citep[][]{Fujimoto_ALPINE_JWST}. For each of the $100$ cube realizations and for each grating of each target, we perform the same \texttt{q3dfit} procedure on the spectra extracted from the ISM and CGM regions. The fitted fluxes in the following discussion are the median value of the 100 realizations with the standard deviation defined as the $1\sigma$ uncertainty. For lines without detection (e.g., lines bluer than [\ion{O}{iii}]$\lambda5007$ for DC-873756), we use the $3\sigma$ upper limit derived from the 100 random cubes integrated over three consecutive wavelength channels, $\sim250-350\,\rm km\,s^{-1}$, around their redshifted line center.

The [\ion{C}{ii}] emission observed by ALMA is extracted in the same ISM and CGM masks and fit in a similar way using \texttt{LMFIT} \citep[][]{Newville_2016LMFIT}.  As we did for the optical lines, we focus on the total flux if more than one Gaussian components are fitted.

\subsection{Line ratio measurements}

We derive line ratios from the median line fluxes of the 100 realizations. The uncertainties of the line ratios are derived by propagating the $1\sigma$ uncertainties of the individual line fluxes. Upper (or lower) limits on the line ratios are derived from the $3\sigma$ upper limits of the line fluxes in the case of their non-detections. 
Following conventions in the literature  \citep[e.g.,][]{Sanders_2024_Te,Sanders_2025}, we use the following abbreviations when referring to line flux ratios in this work:  ${\rm O3}= F_{\rm [\ion{O}{iii}]\lambda5007}/F_{\rm H\beta}$, ${\rm O2}=F_{\rm [\ion{O}{ii}]\lambda3728}/F_{\rm H\beta}$, ${\rm O32} = F_{[\ion{O}{iii}]\lambda5007}/F_{\rm[\ion{O}{ii}]\lambda3728}$, ${\rm R23}=(F_{\rm [\ion{O}{iii}]\lambda\lambda4959,5007}+ F_{\rm [\ion{O}{ii}]\lambda3728})/F_{\rm H\beta}$, ${\rm N2}=F_{[\ion{N}{ii}]\lambda6583}/F_{\rm H\alpha}$, and ${\rm N2O2}=F_{[\ion{N}{ii}]\lambda6583}/F_{\rm[\ion{O}{ii}]\lambda3728}$.

\begin{figure*}[t!]
\centering
\includegraphics[angle=0,width=0.7\textwidth]{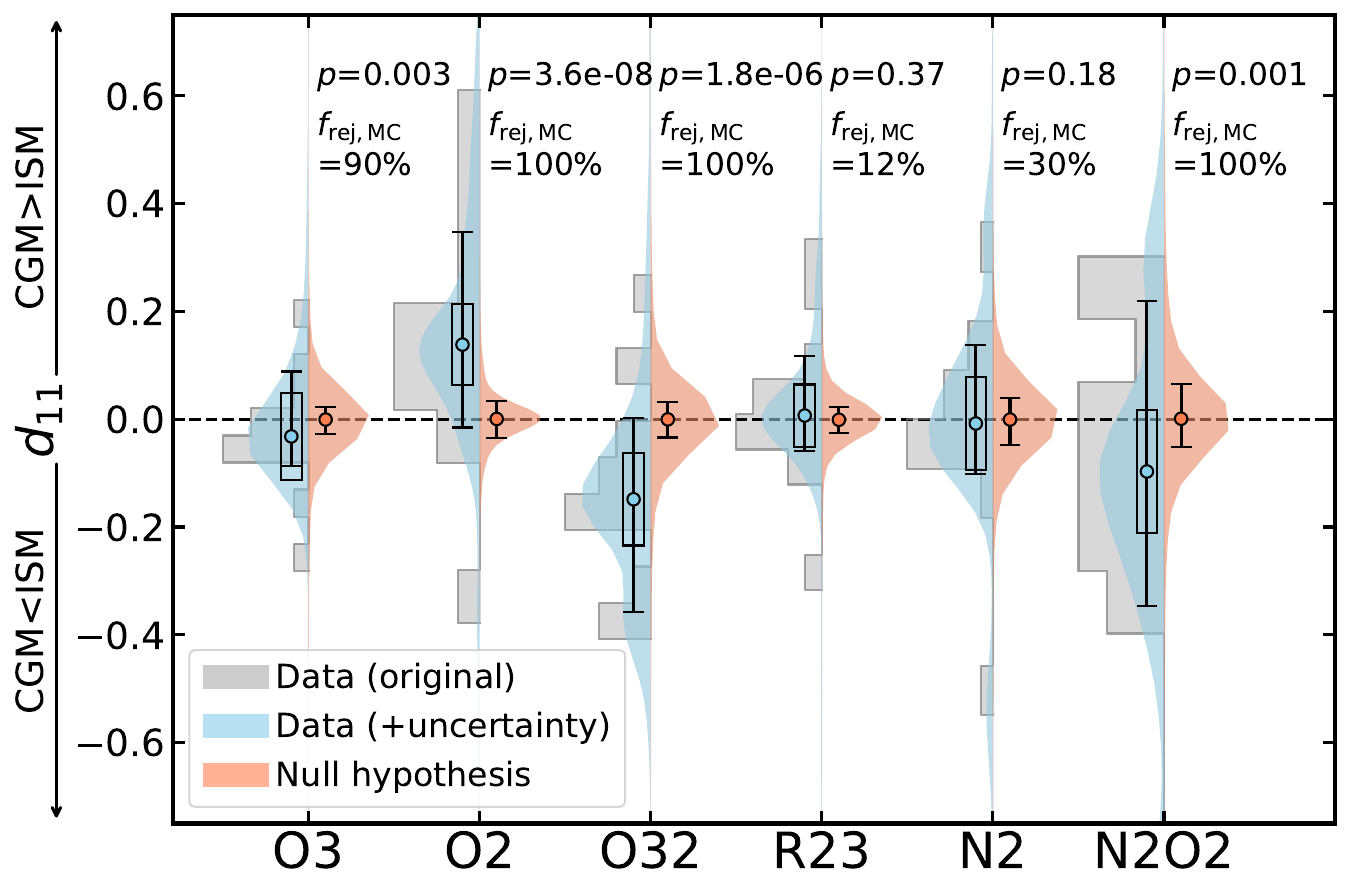}\vspace{-3mm}
\caption{Distribution of data point deviations from the one-to-one line, $d_{11}$, in Fig.~\ref{fig:sample_lineratio}. The gray histograms represent the original data excluding upper and lower limits. The MC sampled distribution of measured $d_{11}$ is shown in light blue and the null hypothesis -- the intrinsic measurement errors centered on the one-to-one line -- is in light red. Circles indicate the median value with error bars marking the 16th and 84th percentile (in corresponding color). Boxes show the interval of $1\sigma_{\rm med}$ standard error around the median values. We mark the p-value of the two-sample KS test and fraction of rejecting null hypothesis based on MC of two-sample KS tests, {\it i.e.} the fraction of MC iterations where the probability of $d_{11}$ around zero is low ($f_{\rm rej, MC} = f_{p<0.05}$, see text). Note that the KS test is preformed on the original data (gray) and the null hypothesis (red) distributions. The blue distributions are {\it only} used for visualization.
}
\label{fig:raio_viol}
\end{figure*}
\section{Results}\label{sec:results}

In this section, we present the measured strong line ratios for the ISM and the CGM. The line ratios can serve as a probe of element abundances or metallicity, but they are also controlled by physical conditions in the gaseous medium, e.g., ionization level, electron temperature ($T_{e}$), and electron density ($n_e$). We discuss the results of the line ratio comparisons in the following section. In this work, we only aim for an analysis of the relative chemical abundance difference between the ISM and CGM instead of a study of absolute metallicity to avoid additional uncertainties brought in the conversion. As stated in Sect. \ref{sec:sample}, we take $Z_{\rm neb}=0.5Z_{\odot}$ as the fiducial metallicity \citep[][]{Faisst_2025b}.

In Fig.~\ref{fig:sample_lineratio}, we compare in each of the panels the measured ISM and CGM line ratios derived following the methods in Sect. \ref{sec:analysis}. We highlight the mergers with different symbols to visually separate them from the non-mergers. The targets with partial transparency in both log(N2) and log(N2O2) panels have ${\rm S/N}_{[\ion{N}{ii}]\lambda6583}<5$. For the following statistical test, we exclude measurements associated with upper or lower limits but still keep the targets with low ${\rm S/N}_{[\ion{N}{ii}]\lambda6583}$ (Sect. \ref{subsec:disc_n2o2} and Appendix \ref{app:nitrogen}).

\subsection{Statistical tests}\label{subsec:statis_test}
To study the line ratio differences between the ISM and the CGM, we adopt two statistical tests. Due to the relatively small sample, we note that the results should be treated with caution. 

Firstly, we conduct the Kendall's $\tau$ test\footnote{For every ordered pair of measurements ($x_i$,$y_i$) and ($x_j$,$y_j$) in the dataset, the test counts the ``concordance" cases where ($x_i>x_j$) \& ($y_i>y_j$) or ($x_i<x_j$) \& ($y_i<y_j$) against the total. The $\tau_{b}$ used in this work is a modified version accounting for ties \citep[][]{KENDALL_1945,kendall1948rank}.} for analyzing the correlations between the ratios in ISM and CGM. The p-value of the Kendall's $\tau$ test indicates the probability of rejecting the null-hypothesis that the two tested samples are independent ({\it i.e.} not correlated). To better take the uncertainties of the data points into consideration, we MC sample over 100 iterations. In each iteration, we randomly sample the data assuming a Gaussian distribution with the width of the measured uncertainty for each data point. In Fig. \ref{fig:sample_lineratio}, we report the $\tau_b$ resulted from the original data, {\it i.e.} no MC sampling. We also show the fraction of iterations where the p-value is $<0.05$, {\it i.e.} the line ratios in the ISM and CGM are correlated, $f_{\rm dep, MC} \equiv f_{p<0.05}=N_{p<0.05}/N_{\rm MC}\times100\%$. Based on the Kendall's $\tau$ test, we find that the ISM and CGM ratios of O32 and O3 are correlated with $f_{\rm dep, MC}>50\%$, especially for O32. The N2 also shows a relatively good correlation between the ISM and the CGM. When excluding outliers, we reach a $f_{\rm dep, MC}=66\%$ for N2.

Secondly, we use the two-sample Kolmogorov-Smirnov (KS) test\footnote{A non-parametric test checks whether two samples are from the same distribution by evaluating the distance between the empirical cumulative distribution functions of the two datasets.} to check the difference between the ISM and the CGM line ratios. Specifically, we calculate the perpendicular distance of each data point to the one-to-one green dashed line on the ISM-CGM plane (Fig. \ref{fig:sample_lineratio}), $d_{11}$. The distribution of measured $d_{11}$ is shown in gray histograms in Fig. \ref{fig:raio_viol}. Our null-hypothesis is that the line ratios are on the one-to-one line, {\it i.e.} equal emission line ratios in ISM and CGM, within the respective uncertainties of the data points. For the null hypothesis distribution, we sample the line ratios 100 times within the uncertainties assuming they land on the one-to-one line and calculate the $d_{11, \rm null}$ (red distribution in Fig. \ref{fig:raio_viol}). The sample KS test is then evaluated between the measured distribution (gray) and the null hypothesis (red) and the corresponding p-values are indicated in Fig. \ref{fig:raio_viol}.
A value $p<0.05$ suggests the rejection of the null-hypothesis, {\it i.e.} the two distributions have different shapes. To better take the uncertainty into account, we adopt the similar MC approach as for the Kendall test by running the KS test 100 times for randomly sampled $d_{11}$. Again, we report the fraction of rejection of the total number of iterations, {\it i.e.} $f_{\rm rej, MC} \equiv f_{p<0.05}=N_{p<0.05}/N_{\rm MC}\times100\%$. Our test shows that the ISM and CGM ratios are different for O2, O32, N2O2, and O3.

To visualize the difference between the measurements and one-to-one line (null-hypothesis), we sample the line ratio 100 times on the ISM-CGM plane from its derived uncertainties. This results in the blue distributions in Fig. \ref{fig:raio_viol}\footnote{This is only used for demonstrating the difference. The KS test has been performed on the original data, {\it i.e.} the gray histograms}. For guiding the comparison, the error bars show the 16-50-84 percentiles of the distributions and the black boxes show the $1\sigma$ standard error on the median with typical value of 0.06~dex\footnote{The standard error on the median is defined as $\sigma_{\rm med}=1.235\sigma/\sqrt{N}$ for a Gaussian distribution, where $\sigma$ is the standard deviation and $N$ is the number of galaxies and merger components in the sample.}.

Combining the Kendall's and KS tests with this visualization, we find the following: 
\begin{enumerate}
    \item The ratio O2 is on average enhanced in the CGM with a $0.14$~dex shift.  
    
    \item The ratio O32 is on average decreased in the CGM with a $-0.15$~dex shift.

    \item The distribution of N2O2 has a relative large range and is skewed towards a deficit in the CGM ($-0.1$~dex shift).

    \item No significant differences are found for the remaining emission line ratios.
    
\end{enumerate}

\subsection{Line ratios versus $M_{\star}$}\label{subsec:lineratio_Mstar}
In nearby star forming galaxies a stronger negative metallicity radial gradient has been reported at higher stellar masses up to $M_{\star}\sim10^{10.5}\,M_{\odot}$ \citep[{\it i.e.} $\sim0.2\,$dex drop over $\sim8$\,kpc,][]{Belfiore_2017,Sanchez_2017}. The ALPINE-CRISTAL-JWST sample ranges in stellar mass from $10^{9.5} - 10^{11.5}\,{\rm M_\odot}$, hence we investigate if such a steepening of the negative metallicity gradient could be observed. To this end, we color-code the symbols in Fig.~\ref{fig:sample_lineratio} by three stellar mass bins of $0.5\,{\rm dex}$ width \citep[$M_{\star}$; derived in][]{mitsuhashi_2024}. From that, we do not see a significant trend of the difference in line ratios between the ISM and CGM with stellar mass. We also note that the distance range studied in \citet{Belfiore_2017} barely reaches into the CGM regime defined in this work. Our results further suggest that the redistribution of metals is statistically similar at all probed stellar masses, which could indicate that the mixing has already been completed in these main-sequence galaxies at the lowest stellar masses.

\section{Discussion}\label{sec:discussion}

Different spatial distributions of gas-phase metallicity between the ISM and CGM reflect the evolution of galaxies. Multiple mechanisms can contribute to the observed diverse metallicity distribution, e.g., inflows, feedback, and merger \citep[e.g.,][]{Rupke_2010,Wang_Enci_2022,Bertemes_2023,Sun_2025_fire}. At lower redshifts, observations often report a negative spatial metallicity gradient (lower metallicity at larger distance) or flat metallicity distributions indicating inside-out enrichment and/or effects by primordial gas accretion onto the star-forming disk \citep[e.g.,][]{Belfiore_2017,Sanchez_2017}. Only a few positive metallicity gradients have been reported, which could indicate outflow redistributing metals \citep[e.g., $\sim10\%$ of nearby spiral galaxies from CALIFA survey,][]{Perez-Montero_2016}. The metallicity gradients for this $z\sim5$ ALPINE-CRISTAL-JWST galaxy sample have been measured in \citet{Fujimoto_ALPINE_JWST,LLee_ALPINE_JWST} out to $\sim3-4\,$kpc. Generally, flat gradients ($\Delta \log(\rm O/H)=0.02\pm0.01\,dex\,kpc^{-1}$) are reported by these studies indicating efficient metal mixing in the ISM.

Complementary to these studies, we characterize the emission line ratios (and therefore metallicity and ionization properties) out to significantly larger radii ($\sim10\,$kpc from the central ionizing source) where the CGM dominates. This is possible thanks to the joint observations with JWST/NIRSpec IFU and deep ALMA observations, which permit such line ratio measurements of the CGM for the ALPINE-CRISTAL-JWST sample. In this section, we first discuss the minimal difference observed between ISM and CGM line ratios indicating a constant chemical abundance (Sect. \ref{subsec:dis_enrich}). For our small sample, our statistical tests suggest there are some provisional difference seen between the line ratios in the ISM and CGM. The complication in performing a simultaneous study of the ISM and CGM is that the physical processes associated with these regions are connected. Furthermore, many of them are degenerate \citep[e.g., ionization and metallicity,][]{Kewley_2002,Dessauges-Zavadsky_2009,Maiolino_2008,Sanchez_2012,Sanders_2024_Te}. Therefore, we discuss the observed line ratio difference in the context of ionization level in Sect. \ref{subsec:disc_ion_abund}, especially for the oxygen line ratios. In Sect. \ref{subsec:disc_n2o2}, we focus on whether the skewed N2O2 distribution (Fig. \ref{fig:raio_viol}) indicates a relative abundance difference. Given the different origins of [\ion{C}{ii}] \citep[][]{Decarli_2025}, we inspect separately the ratio between FIR [\ion{C}{ii}] and H$\alpha$ and discuss its indication on CGM enrichment (Sect.~\ref{subsec:C2dsut}).
Finally, we discuss the properties of the mergers in our sample (Sect.~\ref{subsec:disc_merger}).

\subsection{No significant chemical abundance difference seen between ISM and CGM}\label{subsec:dis_enrich}

Empirical relations have been found between strong line ratios and the gas phase metallicities \citep[e.g.,][]{Maiolino_2008, Steidel_2014,Sanders_2024_Te,Sanders_2025}. Specifically, at the fiducial metallicity of our sample, 0.5$Z_{\odot}$ \citep[][]{Faisst_2025b}, O2 and N2 show a monotonic increase with metallicity while O32 show a monotonic decrease with metallicity. For O3 and R23, the line ratio-$Z_{\rm neb}$ relation is at the turning point around 0.5$Z_{\odot}$, {\it i.e.} not a monotonic relation with metallicity. In Sect. \ref{sec:results}, we find that the R23 and N2 do not show a difference between the ISM and the CGM for our sample. To zeroth order, the difference observed in O3, O2, O32, and N2O2 are not statistically significant, {\it i.e.} less than $3\sigma$, though the KS test suggest a difference (Sect. \ref{subsec:statis_test}). Given that our sample size is relatively small, 18, we could not robustly conclude that we observe a difference in line ratios between ISM and CGM.

We then discuss the indication of the minor difference seen in O3, O2, and O32. Note that the results is derived from a small-sample statistics which should be treated with caution. The KS test indicates that there is a high probability that these three line ratios exhibit a difference between ISM and CGM, especially O2 is higher in the CGM and O3 and O32 are lower in the CGM. If simply assuming that these reflect the gas phase metallicity ({\it i.e.} no difference in ionization between ISM and CGM, see Sect. \ref{subsec:disc_linerat}), the observations imply that the CGM has relatively {\it higher} metallicity, {\it i.e.} higher O/H abundance, than the ISM. Taking the observed ratio difference at face value ($\Delta \log(F_{\rm [\ion{O}{ii}]\lambda3728}/F_{\rm H\beta})_{\rm CGM-ISM}\sim0.14$ and $\Delta \log(F_{\rm [\ion{O}{iii}]\lambda5007}/F_{\rm [\ion{O}{ii}]\lambda3728})_{\rm CGM-ISM}\sim-0.15$) and interpolating the relation from \citet{Maiolino_2008} (or similarly \citealt{Sanders_2024_Te}), we would expect the CGM metallicity (O/H) to be {\it higher} by $\sim0.1-0.2\,$dex. Both \citet{Fujimoto_ALPINE_JWST} and \citet{LLee_ALPINE_JWST} report the metallicity gradient of our sample to be $\sim0.02\,\rm dex\,kpc^{-1}\times10\,kpc$ out to $r\sim4\,$kpc. If we simply extrapolate this relation to the CGM scale probed in this work, $r\sim10\,$kpc, the relation predicts that the CGM metallicity is higher by 0.2~dex than the ISM which would be consistent with such an analysis. However, we note that the cause of this positive metallicity gradient is unclear. Several mechanisms could possibly lead to the distribution of higher metallicity at larger radii. For example, \citet{Sun_2025_fire} find that outflows driven by stellar feedback could potentially trigger a positive metallicity gradient in a small fraction of their galaxies in the FIRE-2 simulation out to $\sim8\,$kpc increasing at most $\sim15\%$. In addition, the accretion of primordial low-metallicity gas can lead to metal-poor star-forming clumps offset, a few kilo-parsecs, from the galaxies' center of mass. The second scenario of inflowing pristine gas is unlikely as these metal-poor clumps in the inner ISM do not sustain for long enough \citep[mixing with the galactic medium within $\lesssim50\,$Myr,][]{Ceverino_2016}, and it does not guarantee that the CGM chemical abundance remain unchanged.  

In summary, our result favor the conclusion that the chemical abundance of the CGM in our sample is enriched to the similar level as in their ISM. If the small number statistic represents the existence of line ratios difference between the ISM and the CGM,  it is unlikely that we observe a positive metallicity gradient. We discuss the alternative scenarios resulting in the observed line ratio difference in Sect. \ref{subsec:disc_linerat}.

\subsection{Line ratio difference: ionization level and relative N/O abundance}\label{subsec:disc_linerat}
 In this section, we further examine two possible explanations for the observed line ratio differences: (i) variation in ionization levels for O3, O2, and O32 (Sect. \ref{subsec:disc_ion_abund}); (ii) variation in relative abundances between nitrogen and oxygen (Sect. \ref{subsec:disc_n2o2}). To assist the discussion, we conduct a simple ionization modeling (Sect. \ref{subsec:cloudy_mod}).

\begin{figure*}[t!]
\centering
\includegraphics[angle=0,width=\textwidth]{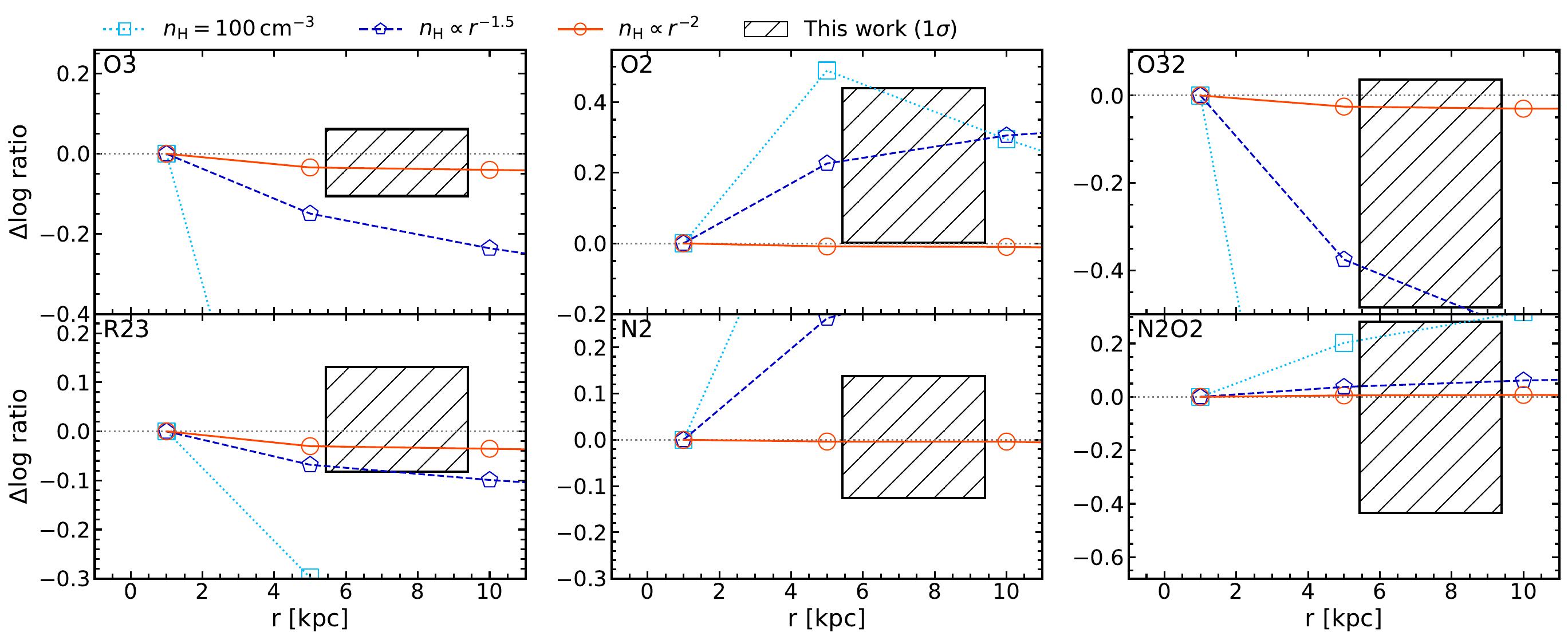}\vspace{-3mm}
\caption{Line ratio differences from \texttt{Cloudy} modeling (assuming no spatial metallicity evolution) measured in thin cloud sheets as a function of distance from the ionizing source (Sect. \ref{subsec:cloudy_mod}). The logarithmic ratio difference at each distance is shown with respect to the line ratio derived as $r=1\,$kpc. We show the modeled radial profiles with $n_{\rm H}=100\,\rm cm^{-3}$ (cyan dotted line with square), $n_{\rm H}=100\times(r/{\rm 1~kpc})^{-2}\,\rm cm^{-3}$ (blue dashed line with pentagon) and $n_{\rm H}=100\times(r/{\rm 1~kpc})^{-1.5}\,\rm cm^{-3}$ (orange solid line with circle). The black hatched box indicates the observed line-ratio range and CGM-size range in our sample (see text).
}
\label{fig:cloudy_mod}
\end{figure*}

\subsubsection{Ionization modeling}\label{subsec:cloudy_mod}
A direct comparison between the chemical properties of ISM and CGM must take into account of different ionization states of these regions as a function of distance from the galaxy's center, {\it i.e.} gas-phase metal abundances is degenerate with ionization levels \citep[e.g,.][]{kewley_2019,hirschmann_2023}. We use a simple \texttt{Cloudy} \citep[][]{Ferland_2017,Chatzikos_2023_cloudy} model to test the impact of a changing distance to the ionization source on the variation of line ratios {\it without} changing the chemical properties and abundance ratios.

{\it Setup}: We provide a brief summary of our \texttt{Cloudy} setup here and describe the details in Appendix \ref{app:cloudy}. The simulation is run using a central ionizing source with the illuminated surface of the cloud being placed, at 1, 5, 10, or 15~kpc to fully cover the CGM size defined for our sample. We test three hydrogen particle number density profiles for the cloud: {\it (i)} $n_{\rm H}=100\,\rm cm^{-3}$ at various distance, 
{\it (ii)} $n_{\rm H}=100\times (r/1\,\rm kpc)^{-1.5}\,\rm cm^{-3}$, and
{\it (iii)} $n_{\rm H}=100\times (r/1\,\rm kpc)^{-2}\,\rm cm^{-3}$. For the element composition of the cloud, we set the relative oxygen abundance to $\rm \log(O/H)=-3.6$ which corresponds to $0.5Z_{\odot}$ of gas phase metallicity. Since the $\rm \log(O/H)$ is fixed in our \texttt{Cloudy} model, we can single-out the change in line ratios as a function of ionization field at a given distance $r$ only.

{\it Result}: The resulting \texttt{Cloudy} models are shown in Fig.~\ref{fig:cloudy_mod} for the strong line ratios focused on in this paper (Sect. \ref{sec:results}). As our goal is to study the line ratio differences between the ISM and CGM, we only show the logarithmic ratio difference with respect to the line ratio at 1~kpc.
We also over-plot the 16th and 84th percentiles of the observed $\log(F_{\rm line1}/F_{\rm line2})_{\rm CGM}-\log(F_{\rm line1}/F_{\rm line2})_{\rm ISM}$ ratio difference derived from the random sampling in Sect.~\ref{sec:results} in black hatched regions. The assumption is that the measured ISM line ratios can be compared to the \texttt{Cloudy} model at $1\,{\rm kpc}$. This assumption is well justified given the size of our ISM region. The width of the box represents the CGM size of our sample, $r_{\rm CGM}$, which is assumed to be the average radius of the circularize CGM mask. Given the large uncertainties and simplification of the modeling, the comparison only ruled out the density profile of constant $n_{\rm H}$.

{\it Notes and caveats:} We emphasize that we do not intend to reproduce the observed physical parameters but only qualitatively study the relative line ratio changes. The primary goal is to visualize the line ratio variation as a function of $r$. It is then physically more intuitive to fix the ionizing parameter, $U$, from the ionizing sources and then decrease it as $\propto r^{-2}$. Our simple setup with placing every cloud at different $r$ ignores the radiative processing through the ISM and CGM. The more realistic scenario will model different layers of clouds for processing and altering the spectrum of ionizing background which is beyond the scope of our analysis.  We note that the absolute value of $n_{\rm H}$ needs to be taken with caution. The $n_{\rm H}=1\,\rm cm^{-3}$ is still higher than typical diffuse CGM \citep[$n_{\rm H}<<0.1\, \rm cm^{-3}$][]{Chen_2024}. The typical PDR has a gas density at $n_{\rm H}\sim10^{2.5-3}\,\rm cm^{-3}$.  The [\ion{C}{ii}] emission has different origins than the optical lines \citep[][]{Decarli_2025} that we do not intend to model with \texttt{Cloudy} in this work. Finally, we comment that we do not consider the contribution from AGN ionization. Our sample is selected to exclude strong AGN \citep[Sect. \ref{sec:sample},][]{Faisst_ALPINE_JWST}. A few are found to show AGN signatures, but the ionization is not dominated by AGN \citep[][]{Ren_AGN}. DC-842313 is the only exception \citep[][]{Solimano_2025}. Since we are modeling the majority of the sample, we exclude photoionization from AGN.

\subsubsection{Abundance variation versus ionization level}\label{subsec:disc_ion_abund}

Figs.~\ref{fig:sample_lineratio} and \ref{fig:raio_viol} suggest that the CGM of the galaxies in our sample has higher O2 and lower O32 than the ISM. In Sect. \ref{subsec:dis_enrich}, we ruled out the possibility that these line ratio differences are due to higher chemical abundance in the CGM than in the ISM. Instead, there is a more intuitive explanation for the observed trend of the line ratios between the ISM and CGM which relates to changing the ionization level \citep[][]{Sanchez_2012} but not the total oxygen abundance, which can be expressed by

\begin{equation}
    \mathrm{\frac{O}{H} = \frac{O^{+}}{H}+\frac{O^{++}}{H}}\text{.}
\end{equation}

Assuming the ionizing radiation is dominated by sources (e.g., young stars) located close to the galaxy center, the CGM will experience less ionizing photons resulting in a lower fraction of ionized ions in higher energy states, e.g., O$^{++}$ \citep[][]{Osterbrock_2006}.
Hence, the observed oxygen line ratio differences between ISM and CGM are due to an increasing of O$^{+}$ and decreasing of O$^{++}$ from ISM to CGM. The observed higher O2 ratio and lower O32 ratio in the CGM both support this argument (Fig.~\ref{fig:cloudy_mod}). The R23, {\it i.e.} a combination of both singly and doubly ionized oxygen lines,  does not show a difference between ISM and CGM (Fig. \ref{fig:raio_viol}) also indicating that there is no clear variance of total oxygen abundance (Sect. \ref{subsec:dis_enrich}).

The comparison to \texttt{Cloudy} (Sect.~\ref{subsec:cloudy_mod}) generally agrees with the above argument in the sense that the difference in ionization level dominates the observed line ratio differences with no clear evidence that oxygen abundance varies between ISM and CGM. As shown clearly in the \texttt{Cloudy} modeled O32, with $n_{\rm H}\propto r^{\alpha},\, \alpha>-2$, the ratio monotonically drops as gaseous cloud moving further from the ionizing source.

In summary, our results indicate that a drop in ionization level in the CGM may be the dominant cause of the observed difference in the oxygen line ratios (such as O2 and O32) and not a change in total oxygen abundance. This would again indicate a constant O/H $\sim10\,$kpc, thus a constant gas-phase metallicity (Sect. \ref{subsec:dis_enrich}). 

\subsubsection{Nitrogen abundances}\label{subsec:disc_n2o2}

The abundance variation of elements from secondary production, such as nitrogen, is more complicated than one of the $\alpha$ elements, such as oxygen, \citep[e.g.,][]{Osterbrock_2006,James_2025}. In general, the N/O abundance ratio stays constant with increasing O/H in the low metallicity regime and increases when secondary nucleosynthesis starts at 12+$\log(\rm O/H)\sim8.3$ \citep[e.g.,][]{Ji_Xihan_2025}. At the metallicity of our sample, 0.5$Z_{\odot}$, we would be in the secondary branch, hence expect an N/O increase with O/H.

The O and N atoms have similar ionization energies ($E_{\rm O^{0}}=13.62\,$eV vs. $E_{\rm N^{0}}=14.53\,$eV), thus N2O2 is expected to be less dependent on the ionization parameter \citep[][]{Zeippen_1982, Fischer_2004, Luridiana_2015_pyneb}. If the $U$ does has an impact, as shown in \citet{Kewley_2002} for the metallicity range of our sample, we would expect a slight increase in N2O2 with decreasing $U$. In our simple \texttt{Cloudy} modeling, Fig. \ref{fig:cloudy_mod}, we indeed find this ratio to increase at larger distances and to be approximately constant for $\alpha=-2$ ({\it i.e.} constant $U$). However, our observation shows a distribution skewed to smaller N2O2 in the CGM compared to the ISM (Fig.~\ref{fig:raio_viol}). This is the opposite of the prediction for the CGM where less ionization is expected (Sect. \ref{sec:results}).

An explanation for this potential discrepancy above could be a nitrogen-oxygen relative abundance difference between the CGM and ISM. This may further suggest two scenarios: {\it (i)} If we assume that N/O $\propto Z$, as mentioned above, the drop of N/O abundance could reflect a decrease in metallicity in the CGM; {\it (ii)} If we assume the secondary nitrogen production has started in the ISM but the additional nitrogen abundance has not yet reached the CGM, the N/O drop indicates a delay in nitrogen enrichment in the CGM. Based on the discussions in Sect. \ref{subsec:dis_enrich} and \ref{subsec:disc_ion_abund}, we rule out scenario {\it (i)}. Scenario {\it (ii)} could be possible given that there might be a $\sim300\,$Myr delay in the nitrogen enrichment. This time is a combination of the nitrogen production and the outflowing from ISM to CGM \citep[e.g.][]{pontinari98,Ginolfi_2020_enrich}. We present this further in Appendix \ref{app:nitrogen}. 

There are several sources of uncertainties to be aware, for example, the dust extinction correction and the S/N of the [\ion{N}{ii}]$\lambda6583$ (${\rm S/N}_{[\ion{N}{ii}]\lambda6583}<5$  detections are marked in Fig. \ref{fig:sample_lineratio} which should be treated with caution). We discuss more on these caveats in Appendix \ref{app:nitrogen}. These further suggest that the difference of the relative nitrogen-oxygen abundance between the ISM and CGM, if any, is not significant.

\begin{figure}[t!]
\centering
\includegraphics[angle=0,width=\columnwidth]{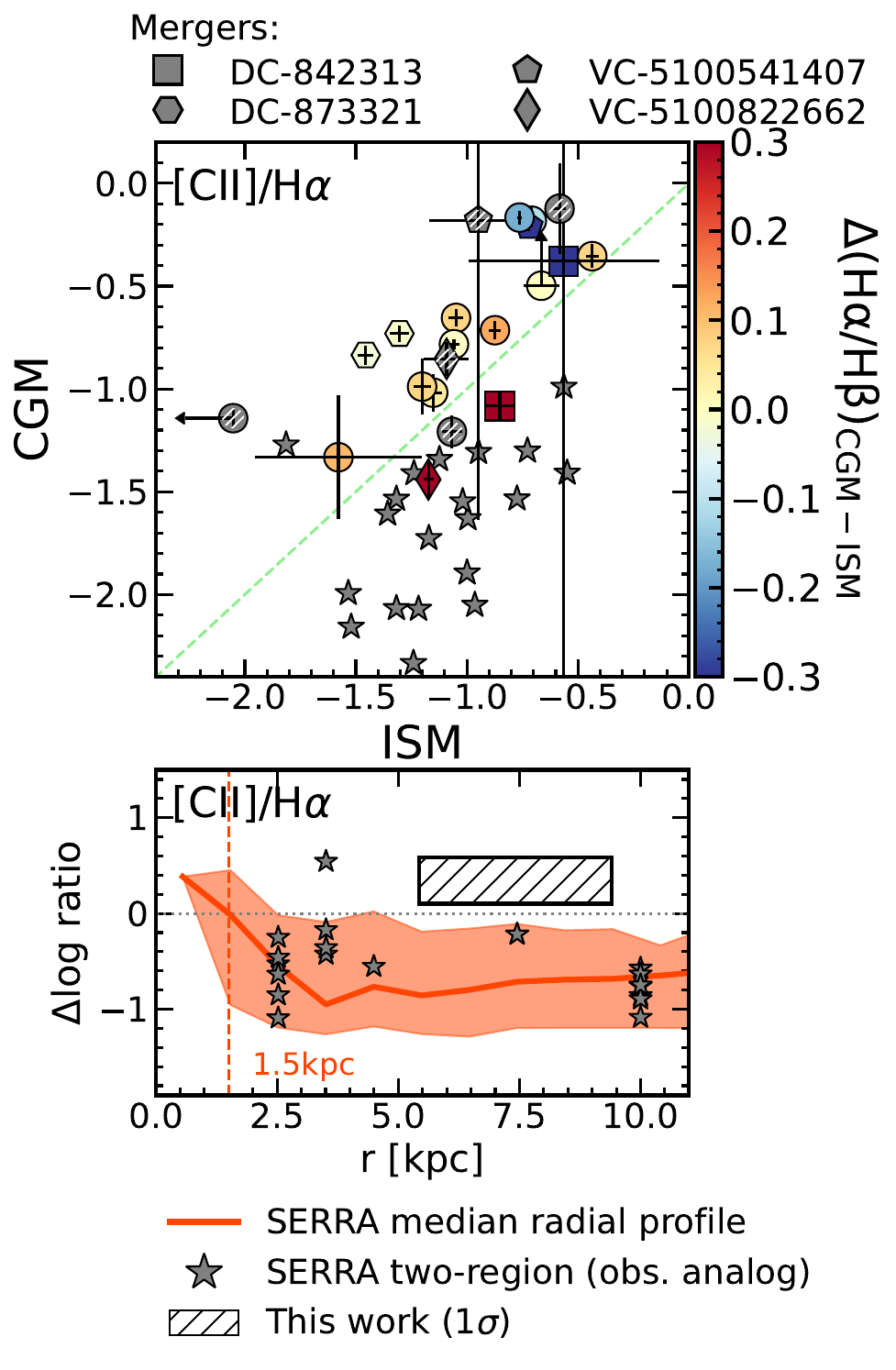}\vspace{-3mm}
\caption{Observed $F_{\rm [\ion{C}{ii}]}/F_{\rm H\alpha}$ and comparison to SERRA simulation. \textit{Upper}: $F_{\rm [\ion{C}{ii}]}/F_{\rm H\alpha}$ ISM versus CGM. The green dashed line marks the one-to-one relation. The data points are color-coded by their $F_{\rm H\alpha}/F_{\rm H\beta}$ difference between ISM and CGM, {\it i.e.} $\log(F_{\rm H\alpha}/F_{\rm H\beta})_{\rm CGM}-\log(F_{\rm H\alpha}/F_{\rm H\beta})_{\rm ISM}$. Positive $\Delta(\rm H\alpha/H\beta)_{\rm CGM-ISM}$ value indicate $F_{\rm H\alpha}/F_{\rm H\beta}$ is large in CGM, {\it i.e.} large dust extinction in CGM than in ISM, and vice versa. The gray color with white hatches indicate that the $\Delta(\rm H\alpha/H\beta)_{\rm CGM-ISM}$ are uncertain (upper limits). We note that the H$\alpha$ line flux used here is corrected for dust extinction. The gray stars with black edges show the results of SERRA galaxies measured in the similar methods as in our observations \citep[see text,][]{Kohandel_2024}.  \textit{Lower}: We show the logarithmic difference in the ratio $F_{\rm [\ion{C}{ii}]}/F_{\rm H\alpha}$ between the CGM and ISM as a black hatched box, similar to Fig.~\ref{fig:cloudy_mod}. Median line ratio difference radial profile of SERRA galaxies are shown in solid red line with shaded region marking the 16th and 84th percentile of the sample. The SERRA profile is normalized to $r=1.5\,$kpc to match the observations. The black stars show the line ratio differences of SERRA galaxies, as in the upper panel. Here, $r$ is the radius where $F_{\rm [\ion{C}{ii}]}$ first drops to $10^{-2}$ of its value at 1.5~kpc. If this threshold is not reached, we set $r=10~$kpc.
}
\label{fig:C2Ha}
\end{figure}

\subsection{[\ion{C}{ii}] emission in the CGM}\label{subsec:C2dsut}

Carbon as one of the most abundant primary metal elements plays an important role in nucleosynthesis. Thanks to the deep ALMA data available for the ALPINE-CRISTAL-JWST sample, we can study carbon emission in the FIR. We note that it is not trivial to directly relate emission from [\ion{C}{ii}] to metallicity \citep[e.g.][]{Lagache_2018}. As shown in \citet{Fudamoto_2025}, [\ion{C}{ii}] traces largely high density neutral gas in star-forming galaxies, e.g., $74-96\%$ of [\ion{C}{ii}] emissions may be related to neutral gas \citep[see also, e.g.,][]{Croxall_2013,Pineda_2013,Levy_2023}. 
In this Section, we investigate the ratio between the FIR fine structure line [\ion{C}{ii}] and H$\alpha$. From Fig.~\ref{fig:C2Ha}, it is clear that the majority of the galaxies have an increased $F_{\rm [\ion{C}{ii}]}/F_{\rm H\alpha}$ ratio in the CGM compared to the ISM. The [\ion{C}{ii}] emission is a major coolant of the gaseous medium while H$\alpha$ can represent the heating of the gas. If we assume that the thermal equilibrium has been reached in the CGM and ISM, our observations may suggest that the CGM gas in general has higher gas density and/or is less ionized \citep[Sect. \ref{subsec:disc_ion_abund}, e.g.,][]{GarciaCarpio_2011,Munoz_2016,RHC_2018a,RHC_2018b}.

We compared our observations to cosmological simulations, specifically the SERRA simulation \citep[zoom-in galaxy formation simulation with on-the-fly radiative transfer,][]{Pallottini_2022,Kohandel_2024,Vallini_2025}.
The SERRA simulation reproduce well the mass-metallicity relation of our sample at $z\sim5$ \citep{Faisst_2025b} as well as samples at higher redshifts \citep{Pallottini_2025_metal}.
From this simulation, we selected 20 galaxies at redshifts ($4.5<z<5.7$), SFR ($>10\,M_{\odot}\,\rm yr^{-1}$), and $M_{\star}$ ($>10^{9.5}\,M_{\odot}$) that resemble the physical properties of the ALPINE-CRISTAL-JWST sample studied here. We follow similar analysis methods as for the real observations.
We derived the $F_{\rm [\ion{C}{ii}]}/F_{\rm H\alpha}$ ratio difference by summing the line fluxes in two spatial regions around the central galaxy; an aperture of $r=3\,$kpc and an annulus with $3<r<15\,$kpc. The results are shown as black stars in Fig. \ref{fig:C2Ha}.
We also present the circular-annulus-extracted radial profiles of $F_{\rm [\ion{C}{ii}]}/F_{\rm H\alpha}$ ratio differences from these 20 matched galaxies in the lower panel of Fig.~\ref{fig:C2Ha}. For a more consistent comparison to the observation, we normalized the profiles at $r=1.5\,$kpc, {\it i.e.} the middle point of the typical ISM size of our sample.
From this comparison, we find that $19$ out of $20$ SERRA galaxies have lower $F_{\rm [\ion{C}{ii}]}/F_{\rm H\alpha}$ ratios in the CGM than in the ISM -- which is in contradiction to our observations. 
This discrepancy could be due to the smaller [\ion{C}{ii}] emitting regions of the SERRA galaxies \citep[e.g. $F_{\rm [\ion{C}{ii}]}$ drops to $10^{-3}$ of the central values at $r=1.5\,$kpc,][]{Pallottini_2022}. The [\ion{C}{ii}] halos of ALPINE-CRISTAL-JWST galaxies have an average half light radius $r_{1/2}^{[\ion{C}{ii}]}=2\,{\rm kpc}$ and do not show a sharp drop up to $r\sim15\,$kpc \citep[][]{Ikeda_2025}. The extension of the [\ion{C}{ii}] emitting region in the SERRA galaxies is never as extended as that observed for the ALPINE-CRISTAL-JWST sample \citep[see also][]{Fujimoto_2019,Faisst_2025b}. 
 We note that forthcoming results from the FIRE-2 high-$z$ simulations with strong feedback show a trend in broad agreement with our findings. In these simulations, the observed extended [\ion{C}{ii}] halos can be reproduced, which can lead to an enhanced $F_{\rm [\ion{C}{ii}]}/F_{\rm H\alpha}$ ratio from the ISM scale out to CGM, closer to that observed in ALPINE-CRISTAL-JWST sample \citep{Liu_2025}.
 This indicates that differences in the feedback implementations between the SERRA and FIRE-2 simulations may be responsible for the different [\ion{C}{ii}] halos extends, and hence the $F_{\rm [\ion{C}{ii}]}/F_{\rm H\alpha}$ flux ratios.

We note that the one outlier SERRA galaxy having elevated $F_{\rm [\ion{C}{ii}]}/F_{\rm H\alpha}$ is due to a satellite in the CGM. This might also the case of our observations where the resolution is not enough to resolve the small satellites. These brighter emissions dominate the flux in the ``smoothed'' data, {\it i.e.} flux-weighted.

The FIR [\ion{C}{ii}] emission is also related to dust. We color-code the data points in Fig.~\ref{fig:C2Ha} by their observed $F_{\rm H\alpha}/F_{\rm H\beta}$ ratio difference between CGM and ISM, {\it i.e.} $\Delta \rm (H\alpha/H\beta)_{CGM-ISM}=\log(F_{\rm H\alpha}/F_{\rm H\beta})_{\rm CGM}-\log(F_{\rm H\alpha}/F_{\rm H\beta})_{\rm ISM}$.
This ratio difference relates to the relative difference in dust extinction between the CGM and the ISM with positive values indicating a relatively higher dust extinction in the CGM. For most galaxies, we do not find a significant difference in dust extinction in their ISM versus their CGM ({\it i.e.} $\Delta \rm (H\alpha/H\beta)_{ISM-CGM}\sim0$). For the three galaxies below the one-to-one line in Fig.~\ref{fig:C2Ha}, two merger components have $\Delta \rm (H\alpha/H\beta)_{CGM-ISM}\gtrsim0.3$. Specifically, DC-842313 (CRISTAL-01) is a complicated system consisting of an AGN, a region of dusty star formation, and kinematic components indicating inflows and outflows \citep[][]{Solimano_2024,Solimano_2025}. This further indicates that merger events could redistribute dust \citep[and also be responsible for the redistribution of metals,][]{Ginolfi_2020}. As for VC-5101244930, the only non-merger below the one-to-one line, it has enhanced H$\alpha$ emissions in the CGM which could be due to satellites or off-center star bursts. In addition, three sources above the one-to-one line show an obvious enhancement of dust extinction in $\Delta \rm (H\alpha/H\beta)_{CGM-ISM}\gtrsim0.1$. 
From our observations, a relation between optical extinction dust and [\ion{C}{ii}] emissions is still elusive. \citet{Veilleux_2025} showed that stellar wind could eject dust to CGM scales of $\sim30\,$kpc. Though the detailed mechanisms are still unclear, our results might show that the feedback and/or merger interactions bring dense gas into CGM in these $4<z<6$ galaxies. A kinematic study is beyond the scope of this work. \citet{Birkin_2025} performed a systematic stack analysis of the [\ion{C}{ii}] emission of ALPINE-CRISTAL-JWST sample where only weak evidence of outflow, {\it i.e.} the presence of broad component, is reported. Nevertheless, we note that a stacking of [\ion{C}{ii}] lines for galaxies with positive and negative $\Delta \rm (H\alpha/H\beta)_{CGM-ISM}$ do not show a clear difference for our sample.

In summary, the higher $F_{\rm [\ion{C}{ii}]}/F_{\rm H\alpha}$ ratio in the CGM than in the ISM found in our sample could be presence of extended [\ion{C}{ii}] emission halos. This may further suggest an effective feedback or merger interaction that brings dense gas to $\sim10\,$kpc. In this process, the dense gas might be ejected where metals may also be associated, but the link is unclear. Deeper observations are required to draw a more robust conclusion.

\subsection{Mergers}\label{subsec:disc_merger}

In our analysis, we separated the ISM and CGM for known mergers \citep[see][for a classification]{Romano_2021}. As shown in Fig.~\ref{fig:sample_lineratio} and \ref{fig:C2Ha}, the ISM and CGM line ratios of mergers have a relatively larger scatter than the non-mergers. For example, the southern component of DC-873321 has an lower O2 ratio in the CGM ($\Delta\log(\rm O2)_{\rm CGM-ISM}\simeq-0.5$), deviating from the rest, and the southern component of VC-5100822662 has a lower N2 in the CGM ($\Delta\log(\rm N2)_{\rm CGM-ISM}\simeq-0.5$). This deviation is expected given that the merger processes can lead to a mixing of metals \citep[e.g.,][]{Rupke_2010}. The merger, DC-873321, could host an AGN which may further complicate the line ratio study \citep[][]{Ren_AGN}.

As mentioned in Sect. \ref{subsec:C2dsut}, DC-842313 (CRISTAL-01), showing a significantly lower $F_{\rm [\ion{C}{ii}]}/F_{\rm H\alpha}$ ratio in its CGM than ISM while having an enhancement in dust extinction in the CGM. The southern component of DC-842313 shows relatively low metallicity as reported in \citet{Solimano_2025} using optical lines (region C01 under their naming scheme, which roughly corresponding to the southern ISM component in this work). Our result further confirms that this southern region has weaker [\ion{C}{ii}] CGM halo. Given that the dusty star forming galaxy and AGN are in the north where a inflow/outflow [\ion{C}{ii}] plume is seen, it may be that the southern halo does not have been enriched yet.
On the other hand, the presence of the AGN in the northern component my provide an increased ionization field that could boost the H$\alpha$ emission, thus reduce the $F_{\rm [\ion{C}{ii}]}/F_{\rm H\alpha}$ ratio. 
We note that the scenario of DC-842313 is complicated where a proto-cluster forming at $z\sim5$ might be seen \citep[][]{Vasan_C01, Solimano_2024,Solimano_2025}.

\section{Conclusions}\label{sec:conclusion}

In this work, we studied the ISM and CGM properties of typical, massive, main-sequence galaxies at $z=4-6$ as part of the ALPINE-CRISTAL-JWST sample. Through pixel-stacking, we are able to probe the CGM out to $\sim10\,{\rm kpc}$ --- farther than other studies on individual galaxies.
For this we used the combined kpc-resolved multi-wavelength imaging and spectroscopic data from HST, JWST, and ALMA, probing the stars, ionized and neutral gas, as well as dust.

By extracting and fitting the spectra in the ISM (stellar continuum) and CGM (expanded ISM region by $\sim2-3$ times), we investigated various commonly used strong emission line ratios (Sect. \ref{sec:results}).
The results are summarized in Figs.~\ref{fig:sample_lineratio} and~\ref{fig:raio_viol}. Our analysis favors the fact that there is no obvious difference of the chemical abundance between the ISM and CGM in our sample (Sect. \ref{subsec:dis_enrich}). This further suggests that the $\sim10\,$kpc halos of ALPINA-CRISTAL-JWST galaxies are enriched to the similar level of their ISM at $z\sim5$ \citep[$Z_{\rm neb}=0.5Z_{\odot}$,][]{Faisst_2025b}. Based on statistical tests (Sect. \ref{subsec:statis_test}), we find that the $F_{\rm [\ion{O}{iii}]\lambda5007}/F_{\rm [\ion{O}{ii}]\lambda3728}$ flux ratio is decreased in the CGM, while the $F_{\rm [\ion{O}{ii}]\lambda3728}/F_{\rm H\beta}$ ratio is elevated in CGM. The $F_{[\ion{N}{ii}]\lambda6583}/F_{\rm[\ion{O}{ii}]\lambda3728}$ flux ratio is skewed towards lower values in the CGM.
We interpret the change in oxygen line ratios as mainly a corollary to the CGM being less ionized (Sect. \ref{subsec:disc_ion_abund}). The other possibility would be a higher metal abundance in the CGM, which is seen in some cases of cosmological simulations but is rather unlikely to be seen consistently through our sample (Sect. \ref{subsec:dis_enrich}).
We also find the possibility of a decreased N/O abundance in the CGM that could explain the observed differences in line ratios. This could be cause by a delay in the redistribution of secondary-produced nitrogen onto the CGM (Sect. \ref{subsec:disc_n2o2}).
Overall, however, the differences between the ISM and CGM are small, thus suggesting a significant chemical enrichment of the CGM at $z\sim5$. This finding is consistent with the relatively flat metallicity gradients at smaller distances of $<3-4\,{\rm kpc}$ \citep[][]{Fujimoto_ALPINE_JWST,LLee_ALPINE_JWST}.

Including the ALMA data, we find that the $F_{\rm [\ion{C}{ii}]}/F_{\rm H\alpha}$ ratio is slightly increased in the CGM,  which may be due to the extended [\ion{C}{ii}] halos seen in this sample.
Interestingly, this ratio increase is not reproduced by the zoom-in simulation SERRA, which also lacks reproducing the extended [\ion{C}{ii}] halos. This suggests that the feedback and/or merger interaction in our sample are more effective in redistributing the dense gas traced by [\ion{C}{ii}] into the CGM. During this processes, the CGM gas may also be metal enriched.

This work is the first comprehensive study of the chemical properties of the ISM and CGM of main-sequence galaxies at $z\sim5$. It will be refined and compared to simulations in more detail in forthcoming papers.

\begin{acknowledgments}
We thank the anonymous referee for their valuable comments and suggestions, which have improved the quality of this manuscript. We also thank Natasha George for suggestions on the writing.
This work is based in part on observations made with the NASA/ESA/CSA \textit{James Webb} Space Telescope. The data were obtained from the Mikulski Archive for Space Telescopes at the Space Telescope Science Institute, which is operated by the Association of Universities for Research in Astronomy, Inc., under NASA contract NAS 5-03127 for JWST. These observations are associated with programs, JWST-GO-01727,  JWST-GO-03045, and JWST-GO-04265. Support for program JWST-GO-03045 was provided by NASA through a grant from the Space Telescope Science Institute, which is operated by the Association of Universities for Research in Astronomy, Inc., under NASA contract NAS 5-03127. W.W. also acknowledges support associated with program JWST-GO-03950.
J.D.S. is supported by JSPS KAKENHI (JP22H01262). 
K.V.G.C. was supported by NASA through the STScI grants JWST-GO-04265 and JWST-GO-03777.
S.F. acknowledges support from the Dunlap Institute, funded through an endowment established by the David Dunlap family and the University of Toronto.
E.d.C. acknowledges support from the Australian Research Council through project DP240100589.
I.D.L. acknowledges funding from the European Research Council (ERC) under the European Union's Horizon 2020 research and innovation program DustOrigin (ERC-2019- StG-851622), from the Belgian Science Policy Office (BELSPO) through the PRODEX project ``JWST/MIRI Science exploitation" (C4000142239) and from the Flemish Fund for Scientific Research (FWO-Vlaanderen) through the research project G0A1523N.
N.G.-V. acknowledges scholarship from ANID BECAS/Doctorado Nacional/2023-21231942
E.I. acknowledges support from ANID MILENIO NCN2024\_112 and ANID FONDECYT Regular 1221846.
J.M. gratefully acknowledges support from ANID MILENIO NCN2024\_112.
M.Relano acknowledges support from project PID2023-150178NB-I00 financed by MCIU/AEI/10.13039/501100011033, and by FEDER, UE. 
K.T. was supported by ALMA ANID grant number 31220026 and by the ANID BASAL project FB210003. 
H.\"U. acknowledges funding by the European Union (ERC APEX, 101164796). Views and opinions expressed are however those of the authors only and do not necessarily reflect those of the European Union or the European Research Council Executive Agency. Neither the European Union nor the granting authority can be held responsible for them.
V. V. acknowledges support from the ANID BASAL project FB210003 and from ANID - MILENIO - NCN2024\_112.
L.V. acknowledges support from the INAF Minigrant "RISE: Resolving the ISM and Star formation in the Epoch of Reionization" (PI: Vallini, Ob. Fu. 1.05.24.07.01).
R.J.A. was supported by FONDECYT grant number 1231718 and by the ANID BASAL project FB210003.
A.N. acknowledges support from the Narodowe Centrum Nauki (NCN), Poland, through the SONATA BIS grant UMO-2020/38/E/ST9/00077.
M.A. is supported by FONDECYT grant number 1252054, and gratefully acknowledges support from ANID Basal Project FB210003 and ANID MILENIO NCN2024\_112.
M.B. acknowledges support from the ANID BASAL project FB210003. This work was supported by the French government through the France 2030 investment plan managed by the National Research Agency (ANR), as part of the Initiative of Excellence of Universit\'e C\^ote d'Azur under reference number ANR-15-IDEX-01.
D.B.S. gratefully acknowledges support from NSF Grant 2407752.
R.H.-C. thanks the Max Planck Society for support under the Partner Group project “The Baryon Cycle in Galaxies” between the Max Planck Institute for Extraterrestrial Physics and the Universidad de Concepción. R.H.-C. also gratefully acknowledges financial support from ANID–MILENIO–NCN2024112 and ANID BASAL FB210003.
L.V. acknowledges support from the INAF Minigrant “RISE: Resolving the ISM and Star formation in the Epoch of Reionization” (Ob. Fu. 1.05.24.07.01).
This paper makes use of the following ALMA data: ADS/JAO.ALMA\#2017.1.00428.L, \#2019.1.00226.S, \#2022.1.01118.S, and \#2021.1.00280.L. ALMA is a partnership of ESO (representing its member states), NSF (USA) and NINS (Japan), together with NRC (Canada), MOST and ASIAA (Taiwan), and KASI (Republic of Korea), in cooperation with the Republic of Chile. The Joint ALMA Observatory is operated by ESO, AUI/NRAO and NAOJ.

\end{acknowledgments}

%

\vspace{5mm}
\facilities{JWST, ALMA}


\software{
\texttt{astropy} \citep{Astropy_2018}; \texttt{Jupyter notebook} \citep[][]{Kluyver_2016_jupyter}; \texttt{matplotlib} \citep[][]{Hunter_2007}; \texttt{SciPy} \citep[][]{Virtanen_2020_scipy}; \texttt{NumPy} \citep[][]{Harris_2020numpy}.
          }



\appendix

\section{\texttt{Cloudy} setup}\label{app:cloudy}
As summarized in Sect. \ref{subsec:cloudy_mod}, we use a central ionizing source and run the \texttt{Cloudy} modeling with clouds at different distances. The spectrum of the central source is modeled using the same SED as that developed in \citet{Veraldi_2025} for ALPINE galaxies, based on the Binary Population and Spectral Synthesis code \citep[BPASS, v2.21][]{Eldridge_2017,Stanway_2018_Bpass}. Specifically, we assume a single starburst with a continuous star formation over 224~Myr \citep[e.g.,][]{Faisst_2020}. The stellar metallicity was set to $0.5Z_{\odot}$ \citep{Faisst_2025b}. We normalized the total UV luminosity to $10^{11}\,L_{\odot}$.

We note that the ionization parameter is defined as
\begin{equation}\label{eq:u}
    U = \frac{Q({\rm H})}{4\pi r^{2} n_{\rm H}c}\text{,}
\end{equation}
where $Q({\rm H})$ is the number of hydrogen ionizing photons per unit time, $r$ is the distance between the ionizing source and the illuminating surface of the cloud, $n_{\rm H}$ is the hydrogen particle number density, and $c$ is the speed of light \citep[][]{Osterbrock_2006}. Hence, our setup of varying $r$ equates to a study of line ratio variation as a function of $U$.
We set the $n_{\rm H}$ to be $100\,\rm cm^{-3}$ at $r=1~$kpc \footnote{With $L_{\rm UV}=10^{11}\,L_{\odot}$, this gives $U\sim5\times10^{-3}$ at the surface of illumination \citep[$Q=\xi_{\rm ion} L_{\rm 1500\AA}$, where  $\xi_{\rm ion}\sim10^{25}\,\rm Hz\,erg^{-1}$,][]{Robertson_2013}.}.
To account for the possibility that the cloud density may be decreasing at larger distance from the galaxy center, we run the simulation with three density profiles (Sect. \ref{subsec:cloudy_mod}).
As $U\propto \left(r^{2} n_{\rm H} \right)^{-1}$, the profile of $n_{\rm H}\propto r^{-2}$ equals to a constant $U$ at the illuminating surface regardless of its distance. 
For each cloud, we run the simulation until a neutral hydrogen column density of $10^{22}~\rm cm^{-2}$ to avoid deep into the photodissociation regions and partially absorption of ionized lines \citep[e.g.][]{Wolfire_2022}. 

\section{Further discussion on N/O abundances}\label{app:nitrogen}

\begin{figure}[t!]
\centering
\includegraphics[angle=0,width=\columnwidth]{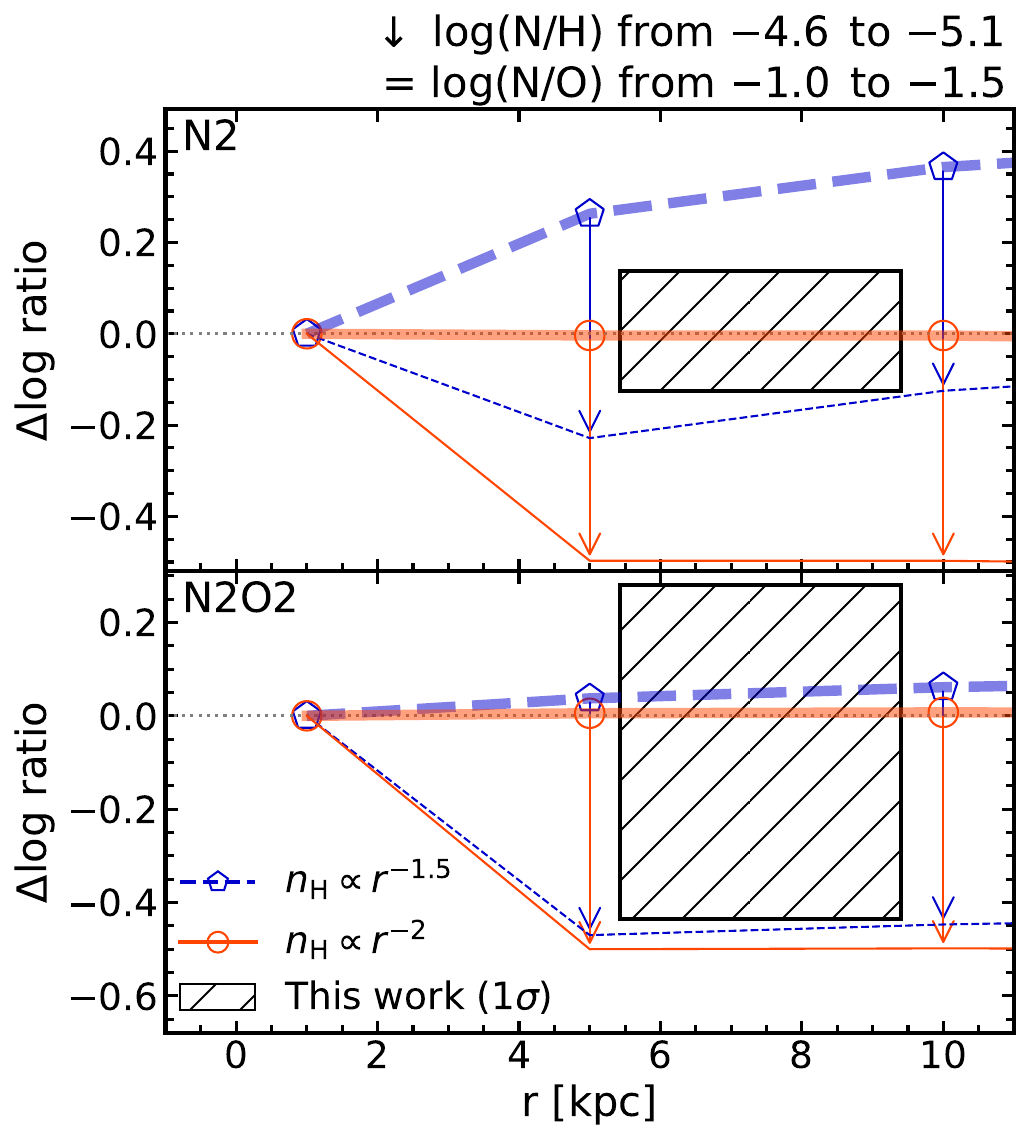}\vspace{-3mm}
\caption{Same as Fig. \ref{fig:cloudy_mod} for N2 and N2O2 showing the \texttt{Cloudy} modeling with varying N/H. The thicker blue dashed line with pentagon (orange solid line with circle) is the same model for $n_{\rm H}=100\times(r/{\rm 1~kpc})^{-1.5}\,\rm cm^{-3}$ ($n_{\rm H}=100\times(r/{\rm 1~kpc})^{-2}\,\rm cm^{-3}$) profile as in Fig. \ref{fig:cloudy_mod}. The arrow in corresponding colors shows the direction of decreasing $\log \rm N/H$. Specifically, it points from the predicted profile with $\log \rm N/H=-4.6\,$dex (thicker line with marker symbols, equivalent to  $\rm \log(N/O)=-1.0\,$dex) to the profile with $\log \rm (N/H)=-5.1\,$dex (thinner line without marker symbols, equivalent to  $\rm \log(N/O)=-1.5\,$dex). The $\log \rm (N/H)=-4.6\,$dex is fixed for cloud at $r=1\,$kpc.
}
\label{fig:cloudy_varyN}
\end{figure}

In Sect. \ref{subsec:disc_n2o2}, our discussion shows that the observed N2O2 indicates a relative nitrogen-oxygen abundance difference between the CGM and ISM. To additionally test the relation between N2O2 and N/O abundance in our \texttt{Cloudy} simulation (Sect. \ref{subsec:cloudy_mod} and Appendix \ref{app:cloudy}), we vary the $\rm \log(N/H)$ for clouds at $r>1\,$kpc but keep the O/H, $Z_{\rm neb}$, fixed such that the net effect equals to a varying of $\rm \log(N/O)$. We test a change in the nitrogen abundance between $\rm \log(N/H)=-4.6$ and $-5.1$  \citep[this corresponds to $\rm \log(N/O)=-1.0$ and $-1.5$, respectively, as expected from observations,][]{Ji_Xihan_2025} between the ISM and CGM, respectively.
To this end, we fix $\rm \log(N/H)=-4.6$ for a cloud at $r=1\,$kpc. We visualize the changes in the profile from a higher ($\rm \log(N/O)=-1.0$; line with markers) to a lower ($\rm \log(N/O)=-1.5$; line without marker) N/O abundance by arrows in Fig.~\ref{fig:cloudy_varyN}. The observations agree more with a shallow drop in N/O abundance from the ISM to the CGM. This simple zeroth-order modeling may indicate that there is a variation in N$^{+}$ (or N$^{+}$/O$^{+}$) between ISM and CGM (Sect. \ref{subsec:disc_n2o2}).

Under the assumption that the observed line ratio difference is resulted from abundance difference, we can calculate the difference following \citet{Pagel_1992} who provided a relation between the relative abundance of nitrogen and oxygen ions and $(F_{[\ion{N}{ii}]\lambda6548}+F_{[\ion{N}{ii}]\lambda6583})/F_{\rm[\ion{O}{ii}]\lambda3728}$ \citep[see also,][]{Belfiore_2017},
\begin{align}\label{eq:N2O2_ab}
    \log({\rm N/O}) =~& \log({\rm \frac{4}{3}N2O2})+0.307-0.02\log t_{\rm N^{+}}\notag\\
    &-0.726 t_{\rm \rm N^{+}}^{-1} + {\rm \log(ICF(N^{+}/O^{+}))}\text{,}
\end{align}
where $t_{\rm \rm N^{+}}$ is the electron temperature of [\ion{N}{ii}] emitting nebula in the unit of $10^{4}\,$K. In this equation we added a factor of 4/3 because our definition of N2O2 does not use $F_{[\ion{N}{ii}]\lambda6548}$ ($F_{[\ion{N}{ii}]\lambda6548}=1/3F_{[\ion{N}{ii}]\lambda6583}$).
Note that an ionization correction factor  (ICF)\footnote{$\rm N/O =  ICF(N^{+}/O^{+})\times (N^{+}/O^{+}$).} needs to be applied for an exact equality. The $\rm ICF(N^{+}/O^{+})$ varies with metallicity and oxygen ion abundance and could cause a $0.2\,$dex uncertainty in abundance \citep[e.g.,][]{Izotov_2006}. The temperature $t_{\rm \rm N^{+}}$ is assumed to be similar to $t_{\rm \rm O^{+}}$. The latter can be calibrated using R23 \citep[e.g.,][]{Thurston_1996},
\begin{align}
    t_{\rm N^{+}} \sim t_{\rm O^{+}} =~& 6065 + 1600(\log \rm R23)+ 1878(\log \rm R23)^{2} \\
    &+1803(\log \rm R23)^{3}\notag\text{.}
\end{align}
Because our observations suggest that the R23 ratio stays unchanged for the ISM and CGM, we assume that also $t_{\rm N^{+}}$ is constant. 
With this, we can estimate the abundance difference $\Delta \rm \log( N^{+}/O^{+})_{\rm CGM - ISM}\sim-0.2$ to $-0.4$ using Eq. \ref{eq:N2O2_ab}. 
In addition, we can apply an approximate ICF \citep[e.g., $\log(\rm N/O)=\log(N^{+}/O^{+})+0.1$,][]{Amayo_2021} resulting in a $\Delta \rm \log(N/O)_{\rm CGM - ISM}\sim-0.1$ to $-0.3$. Similarly, 
\citet{Strom_2017} where a relation between N/O and N2O2 is reported based on observations of $z\approx2$ galaxies,
\begin{equation}\label{eq:S17N2O2}
    \log(\rm N/O) = 0.65\times \log({\rm N2O2}) -0.57\text{.}
\end{equation}
Using this calibration we find $\Delta \rm \log(N/O)_{\rm CGM - ISM}\lesssim-0.2$ for our sample, consistent with the result above.

As briefly mentioned in Sect. \ref{subsec:disc_n2o2}, the N/O abundance difference between ISM and CGM could be due to a delay in nitrogen enrichment. AGB stars would be the dominant source of nitrogen enrichment \citep[typically at intermediate masses of $4-7\,{\rm M_\odot}$;][]{Nomoto_2013}. If we simply assume that the outflow distributing newly produced nitrogen has a velocity of $v\sim200\rm \,km\,s^{-1}$ \citep{Ginolfi_2020_enrich}, it would take $\sim100\,$Myr for CGM to be enriched out to distances of $10\,{\rm kpc}$. Given that and the typical time for significant nitrogen production \citep[$\gtrsim200\,{\rm Myr}$;][]{pontinari98}, there may be a delay of $>300\,{\rm Myrs}$ for the nitrogen enrichment of the CGM from ejecta from the ISM.
Such a difference in the N/O abundance ratio would change the emission line ratios in a way that is consistent with observations (see \texttt{Cloudy} modeling in Fig.~\ref{fig:cloudy_varyN}). However, still ionization plays an important role as shown by the O32 line ratio which is independent of the N/O abundance. Wolf-Rayet (WR) stars with strong stellar winds can also enhance the nitrogen abundance on a short time scale of $\sim3-5\,$Myr \citep[][]{Crowther_2007}, but the evidence of appearance of WR stars in our sample requires more investigation.

Finally, we discuss the uncertainties that could impact the discussion on nitrogen-oxygen relative abundance difference. The dust extinction could affect N2O2 as $[\ion{O}{ii}]\lambda3728$ is more obscured than $[\ion{N}{ii}]\lambda6583$. 
The assumed intrinsic H$\alpha$/H$\beta$ may not be representative of the $n_{e}$ and $T_{e}$ of the CGM gas \citep[][]{Faisst_ALPINE_JWST}. Furthermore, we use H$\beta$ on G395M if it is covered by both of the gratings (Sect. \ref{subsec:ana_specfit}). The 6$\%$ higher flux on G395M than on G235M will impact the extinction correction resulting in an overestimation of the N2O2 by $\sim0.07~$dex which is comparable to the $1\sigma$ uncertainty (0.06~dex, Sect. \ref{subsec:statis_test}). This flux difference impacts more on ISM than CGM indicating the possibility that the N2O2 may not show a difference between the ISM and CGM. Additionally, a low S/N measurement in the [\ion{N}{ii}]$\lambda6583$ line may cause significant uncertainties in the N/O abundance determination. We test this by fitting simulated [\ion{N}{ii}]$\lambda\lambda6548,6583$ and H$\alpha$ line pairs. To this end, we add noise to set S/N$_{[\ion{N}{ii}]\lambda6583}$ ratios of 3, 5, 10, and, 20. For each of these S/N, we refit the line pairs to obtain their fluxes. We find that at S/N$_{[\ion{N}{ii}]\lambda6583}<5$, the fitted $F_{[\ion{N}{ii}]\lambda6583}$ may underestimate the intrinsic value by a factor of $\sim2$ (=0.3$\,$dex). In Fig.~\ref{fig:sample_lineratio} (panel log(N2O2)), we therefore mark the sources with CGM S/N$_{[\ion{N}{ii}]\lambda6583}<5$ in transparent color \footnote{The S/N is not an issue in the ISM component.}. While the observed N2O2 is still skewed to lower values in the CGM, this test shows that the N/O abundance change may not be as significant as in the case where these low S/N sources are included. Lastly, the N2 ratio alone should also be affected by the deficit of nitrogen. Since we do not observe a difference in N2 between CGM and ISM (Fig.~\ref{fig:cloudy_mod}), the explanation of a delay in nitrogen production for the drop of N2O2 is still elusive.





\bibliography{references}{}
\bibliographystyle{aasjournalv7}


\allauthors
\end{CJK*}
\end{document}